\documentclass[acmsmall]{acmart}

\setcopyright{cc}
\setcctype{by}
\acmDOI{10.1145/3797099}
\acmYear{2026}
\acmJournal{PACMSE}
\acmVolume{3}
\acmNumber{FSE}
\acmArticle{FSE071}
\acmMonth{7}
\received{2025-09-10}
\received[accepted]{2025-12-22}

\usepackage{adjustbox}
\usepackage{amsmath,amssymb,amsfonts}
\usepackage[linesnumbered,ruled,vlined]{algorithm2e}
\usepackage{setspace}
\usepackage{algorithmic}
\usepackage{float}
\usepackage{graphicx}
\usepackage{textcomp}
\usepackage{xcolor}
\usepackage{multirow}
\usepackage{mdframed}
\usepackage{balance}
\usepackage{booktabs}
\usepackage{tcolorbox}
\usepackage{makecell}
\usepackage{threeparttable}
\usepackage{colortbl}
\usepackage{subfigure}
\usepackage{hhline}
\usepackage{pifont}
\usepackage{listings}
\usepackage{enumitem}
\usepackage{caption}

\usepackage{wrapfig}

\definecolor{ggray}{HTML}{eff0f0}
\definecolor{gggray}{HTML}{E8E8E8}
\definecolor{ggggray}{HTML}{BEBEBE}

\newcommand{\sota}{state-of-the-art}

\newcommand{\ie}{\textit{i.e.,}}
\newcommand{\eg}{\textit{e.g.,}}

\newcommand{\agentless}{\textsc{Agentless}}
\newcommand{\autocoder}{\textsc{AutoCodeRover}}

\newcommand{\deepseek}{Deepseek-Coder-6.7b-instruct}
\newcommand{\llama}{Llama-3-8B-Instruct}
\newcommand{\chatglm}{ChatGLM2-6b}
\newcommand{\DP}{Deepseek-Coder-33b-instruct}

\newcommand{\ourtool}{\textsc{TransAGENT}}
\newcommand{\unitrans}{\textsc{UniTrans}}
\newcommand{\transmap}{\textsc{TransMap}}
\newcommand{\transcoder}{\textsc{TransCoder}}
\newcommand{\sda}{\textsc{SDA-Trans}}
\newcommand{\batfix}{\textsc{BatFix}}

\newcommand{\Abbourtool}{\textsc{TA}}
\newcommand{\Abbunitrans}{\textsc{UT}}

\usepackage[utf8]{inputenc}
\usepackage[english]{babel}

\newcounter{finding}

\newcommand{\distance}{5pt}
\begin{document}

\title{TransAgent: Enhancing LLM-Based Code Translation via Fine-Grained Execution Alignment}

\author{Zhiqiang Yuan}
\orcid{0000-0002-6497-9380}
\affiliation{%
  \institution{Fudan University}
  \city{Shanghai}
  \country{China}
}
\email{zhiqiangyuan23@m.fudan.edu.cn}

\author{Weitong Chen}
\orcid{0009-0003-3351-9650}
\affiliation{%
  \institution{Fudan University}
  \city{Shanghai}
  \country{China}
}
\email{25213050003@m.fudan.edu.cn}

\author{Hanlin Wang}
\orcid{0009-0008-2374-1246}
\affiliation{%
  \institution{Fudan University}
  \city{Shanghai}
  \country{China}
}
\email{23210240303@m.fudan.edu.cn}

\author{Xin Peng}
\orcid{0000-0003-3376-2581}
\authornote{Corresponding author: Xin Peng (pengxin@fudan.edu.cn)}
\affiliation{%
  \institution{Fudan University}
  \city{Shanghai}
  \country{China}
}
\email{pengxin@fudan.edu.cn}

\author{Zhenpeng Chen}
\orcid{0000-0002-4765-1893}
\affiliation{%
  \institution{Tsinghua University}
  \city{Beijing}
  \country{China}
}
\email{zpchen@tsinghua.edu.cn}

\author{Yiling Lou}
\orcid{0000-0001-7814-0693}
\affiliation{%
  \institution{Fudan University}
  \city{Shanghai}
  \country{China}
}
\email{yilinglou@fudan.edu.cn}

\begin{abstract}
Code translation transforms code between programming languages while preserving functionality, which is critical in software development and maintenance. 
While traditional learning-based code translation methods have limited effectiveness due to the lack of sufficient parallel training data, Large Language Models (LLMs) have recently advanced this field with their strong code generation and comprehension capabilities. However, code translated by LLMs still suffers from diverse quality issues, such as syntax and semantic errors. In this work, we propose \ourtool{}, a novel multi-agent system that eliminates the errors during LLM-based code translation. The main insight of \ourtool{} is to localize error-prone code blocks via fine-grained execution alignment between source and target code. 
We evaluate \ourtool{} on a newly constructed benchmark of recent programming tasks to mitigate data leakage. 
\ourtool{} outperforms the latest \unitrans{} by up to 33.3\% in translation accuracy and achieves an average improvement of 56.7\% over \agentless{} in program repair performance. We also conduct an ablation study and evaluate \ourtool{} across different LLMs, demonstrating its effectiveness and strong generalizability.

\end{abstract}



\maketitle

\section{Introduction}
Code translation aims at transforming the code from one programming language (\ie{} source program) to another (\ie{} target program) while still preserving the same functionality.  
Code translation is prevalent in software development and maintenance, given the needs of performance optimization~\cite{9678878, 7273801}, system modernization~\cite{DBLP:journals/corr/abs-2004-10724, DBLP:conf/euromicro/HaugelandNSC21}, or technology transitions. While manually performing code translation can be time-consuming and error-prone, various automated code translation techniques (\ie{} transpilers)~\cite{DBLP:conf/nips/RoziereLCL20, DBLP:conf/iclr/RoziereZCHSL22, DBLP:conf/iclr/SzafraniecRLLCS23,unitrans} have been proposed.

Traditional rule-based code translation relies on manually written transformation rules to convert source code into target code~\cite{JavaToCSharp, cxgo, sharpen}. However, crafting these rules requires extensive effort, and the resulting target programs often suffer from poor readability and usability~\cite{DBLP:conf/nips/RoziereLCL20}. 
To address this issue, a series of learning-based code translation methods has been proposed to enhance translation quality~\cite{DBLP:conf/nips/RoziereLCL20, DBLP:conf/iclr/RoziereZCHSL22, DBLP:conf/iclr/SzafraniecRLLCS23}. 
These methods train models on large amounts of parallel data (\ie{} paired source and target programs), enabling the models to learn translation patterns and mappings between different languages. However, high-quality parallel data for training is often scarce in practice~\cite{DBLP:conf/iclr/RoziereZCHSL22, DBLP:conf/eacl/AhmadCRC23, DBLP:conf/emnlp/XieNFR23}, and the process of model training is also very time-consuming. For example, training the \transcoder{} model requires 32 V100 GPUs for 12 days~\cite{DBLP:conf/nips/RoziereLCL20}.

Recent advancements in large language models (LLMs) further enhance learning-based code translation. The translation paradigm has also shifted from ``train-then-translate'' to ``translate-then-fix'', where LLMs first translate the source into the target program and then fix translation errors~\cite{DBLP:conf/icse/PanIKSWMSPSJ24}. Translation errors in target programs fall into two categories: \textit{syntax errors}, which violate the grammar of the target language, and \textit{semantic errors}, where the program produces outputs that differ from the source program for the same input. Effective use of LLMs for error repair requires precise localization and concrete guidance~\cite{DBLP:conf/icse/XiaWZ23, AutoCodeRover, Agentless, DBLP:conf/aaai/JoshiSG0VR23}. Syntax errors can be detected by compilers, but the suggested fixes are often not straightforward. Semantic errors are difficult to locate, as they manifest only as output discrepancies without causing compilation or runtime failures.

However, existing methods rely solely on end-to-end LLM reasoning to fix translation errors, which limits their effectiveness~\cite{DBLP:conf/apsec/TaoYGS24, unitrans, DBLP:conf/icse/PanIKSWMSPSJ24, DBLP:journals/corr/abs-2412-14234}. For example, \unitrans{}, the latest LLM-based translation approach, leverages function-level test cases to guide LLMs in repairing semantic errors in the target program. 
Since LLMs cannot accurately access the dynamic runtime behavior of the target program~\cite{chen2024reasoning}, this prevents the detection of fine-grained logical discrepancies and the pinpointing of the exact location to repair.
Consequently, \unitrans{} produces only a marginal accuracy gain in the error-fixing stage, increasing from 31.25\% to 31.90\% on the Java-to-Python task with Llama-7B~\cite{unitrans}.

To enhance LLM in code translation, particularly for fixing syntax and semantic errors in the target program, we propose \ourtool{}, an LLM-based multi-agent system. 
\textit{The main insight of \ourtool{} is to pinpoint the error locations in the target program and provide specific fix suggestions, thus reducing the difficulty of error fixing for LLMs.}
\ourtool{} employs four collaborative agents: \textit{Initial Code Translator}, \textit{Syntax Error Fixer}, \textit{Code Aligner}, and \textit{Semantic Error Fixer}. 
First, \textit{Initial Code Translator} generates test cases and an initial target program for the given source program; Second, \textit{Syntax Error Fixer} captures compiler error messages, converts them into more specific fix suggestions, and leverages LLM to fix the syntax errors; Third, \textit{Code Aligner} divides the source program into blocks using the control flow graph (CFG) and then maps each block to the target program with LLM; Lastly, based on the mapped blocks between the source and target program, \textit{Semantic Error Fixer}  localizes the error block in the target program which exhibits different runtime behaviors from its aligned block in the source program, and then uses LLMs to specifically fix the error block with the observed runtime difference.

Pinpointing semantic errors in the target program is challenging, as it requires precise knowledge of the runtime values of variables and how they deviate from their expected outputs. Intuitively, this challenge can be addressed by comparing the intermediate runtime states of the source and target programs, much like developers inspect program states to locate errors by debugging.
However, identifying the correspondence between the source and target programs is challenging, as structural differences frequently arise between their implementations.
Existing methods like \transmap{}~\cite{transMap} rely solely on LLMs for statement-level alignment, but this approach is fragile; for example, if a source line translates into multiple target lines or lines are shifted, LLMs often default to naive sequential mapping, resulting in misalignments~\cite{transMap}.
To address the mapping issue, \ourtool{} introduces the \textit{Code Aligner}, which leverages CFG-based decomposition to partition source code into atomic blocks and align them with the target program using LLMs. 
This block-level alignment is more robust than statement-level alignment; for instance, it groups sequential lines into a single block and maps it to the target program, effectively resolving issues such as line shifts.

Based on the aligned blocks, the \textit{Semantic Error Fixer} simultaneously executes the source and target programs, compares the runtime states of variables within the mapped blocks, detects divergences, pinpoints the faulty block, and performs targeted repairs.
Unlike traditional fault-localization~\cite{DBLP:conf/uss/BlazytkoSAAFWH20, DBLP:conf/issta/JinO13} and repair methods~\cite{DBLP:conf/aaai/GuptaPKS17, DBLP:conf/uss/BlazytkoSAAFWH20} that operate on a single program, the \textit{Semantic Error Fixer} leverages cross-program execution comparison to guide precise corrections. Moreover, unlike coarse-grained approaches such as file- or function-level localization~\cite{AutoCodeRover, Agentless}, \ourtool{} performs fine-grained, block-level localization, where each block consists of several lines of code, to enhance LLM accuracy in correcting translation errors.

\begin{figure*}[t]
    \centering
    \includegraphics[width=0.95\textwidth]{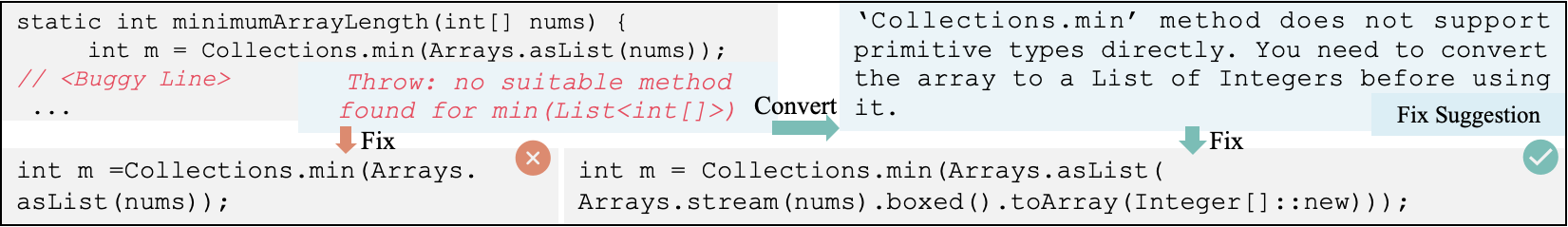}
    \caption{Example of Fixing Syntax Errors in Target Java Program}
    \label{fig:syntax_fix}
\end{figure*}

To evaluate \ourtool{}, we first construct a new benchmark from recent programming tasks to mitigate the potential data leakage issue. 
We then evaluate the translation performance of \ourtool{}, which achieves 87.1\% higher accuracy than the learning-based \transcoder{} and 33.3\% higher than the LLM-based \unitrans{}. An ablation study reveals that \textit{Syntax Error Fixer} and \textit{Semantic Error Fixer} substantially enhance translation performance, surpassing the error-fixing strategy of \unitrans{}.
\ourtool{} also demonstrates superior repair accuracy, exceeding \agentless{} by 56.7\% and \ourtool{}\textsubscript{TM} (with the \transmap{} strategy) by 14.5\%. Finally, \ourtool{} exhibits strong generalization to different LLMs.

We summarize our contributions as follows:
\begin{itemize}
\item \textbf{A Novel LLM-based Code Translation Technique.} We propose \ourtool{}, an LLM-based  multi-agent system for  fixing the syntax and semantics errors in LLM-based code translation. 

\item \textbf{A Novel Code Mapping Strategy.} We design a code mapping strategy (\ie{} \textit{Code Aligner}) upon the synergy between control flow analysis and LLMs. 

\item \textbf{A New Code Translation Benchmark.} We construct a new code translation benchmark, which is constructed from the recent programming tasks, so as to mitigate the data leakage issue when evaluating LLM-based code translation techniques. 

\item \textbf{Comprehensive Evaluation.} We systematically evaluate \ourtool{} across various perspectives, including the overall translation effectiveness, the ablation study of each agent, repair accuracy, and generalization. 
\end{itemize}

\section{Motivating Example}\label{sec:motivate}

We illustrate the error-fixing challenges in the latest LLM-based code translation technique, \unitrans{}~\cite{unitrans} via two examples. Here, we use the version of \unitrans{} built upon the backbone LLM \deepseek{}~\cite{deepseek}. 
The first example demonstrates a failure to fix a \textit{syntax error}, while the second shows a failure to resolve a \textit{semantic error}.

\textbf{Challenges in fixing syntax errors solely with the error messages thrown by the compiler/interpreter.}
\unitrans{} relies on compiler error messages to fix syntax errors, but these messages are often vague and lack specific repair guidance, making it difficult for LLMs to resolve issues effectively. For instance, in Figure~\ref{fig:syntax_fix}, the Java program \texttt{minimumArrayLength}~\cite{minimumArrayLength}, translated from Python, encounters a compilation error: \texttt{no suitable method found for min(List<int[]>)}. The error message fails to explain why the method call is invalid or how to fix it, providing limited hints for the LLMs to generate correct patches.  To enhance the ability of LLMs to correct syntax errors, it is helpful to convert compiler error messages into clearer, more specific fix suggestions. For example, rephrasing the message to indicate that \texttt{Collections.min()} requires a \texttt{List} of objects rather than primitives would better guide the LLM in resolving the issue.

\begin{figure*}[t]
    \centering
    \includegraphics[width=0.95\textwidth]{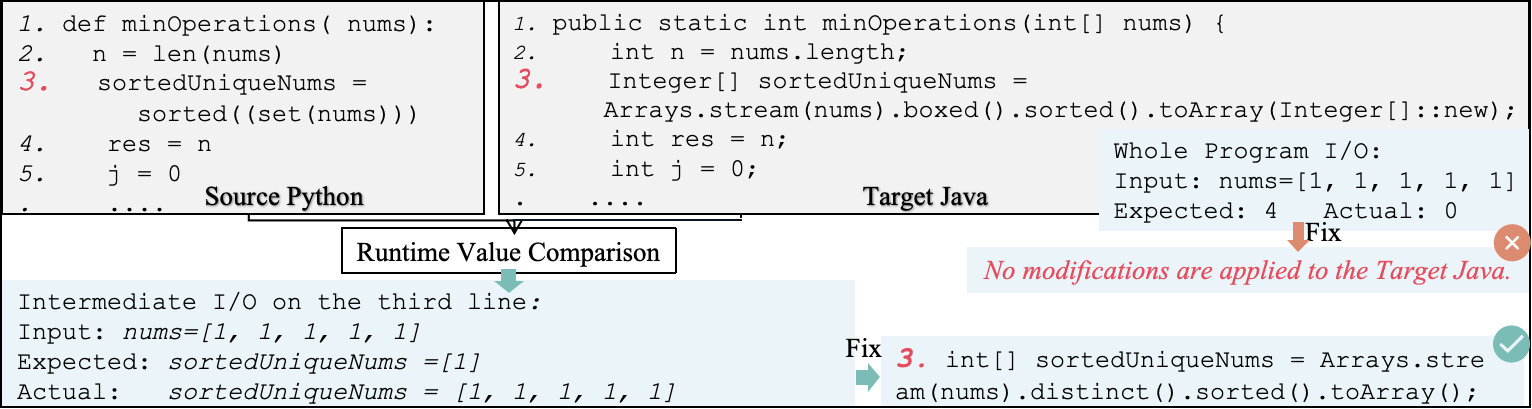}
    \caption{Example of Fixing Semantic Errors in Target Java Program}
    \label{fig:Sem_fix}
\end{figure*}

\textbf{Challenges in fixing semantic errors solely with whole program input and output.}
\unitrans{} relies on test inputs and outputs of the entire program to address semantic errors, but this approach is often too complex for LLMs. Reasoning along full execution paths, especially with multiple logical branches, exceeds the capabilities of LLMs, as noted in recent studies~\cite{chen2024reasoning}. For example, in Figure~\ref{fig:Sem_fix}, the Java program \texttt{minOperations}~\cite{minOperations} translated by \unitrans{} from Python misses a key deduplication operation in Line \texttt{3} of the source Python program. This omission causes the Java program to return an incorrect output (\texttt{0} instead of \texttt{4}). Despite using test inputs and outputs, \unitrans{} fails to correct this subtle logic error due to program complexity. Improving semantic error correction could involve decomposing the problem by localizing fine-grained error locations (\eg{} specific statements or blocks) and using relevant runtime values to fix them. For instance, comparing variable runtime values between the Python and Java programs could pinpoint the error to Line \texttt{3} in the Java program, providing more focused hints for LLMs to resolve the issue.


\section{Approach}
\begin{figure*}[t]
    \centering
    \includegraphics[width=0.8\textwidth]{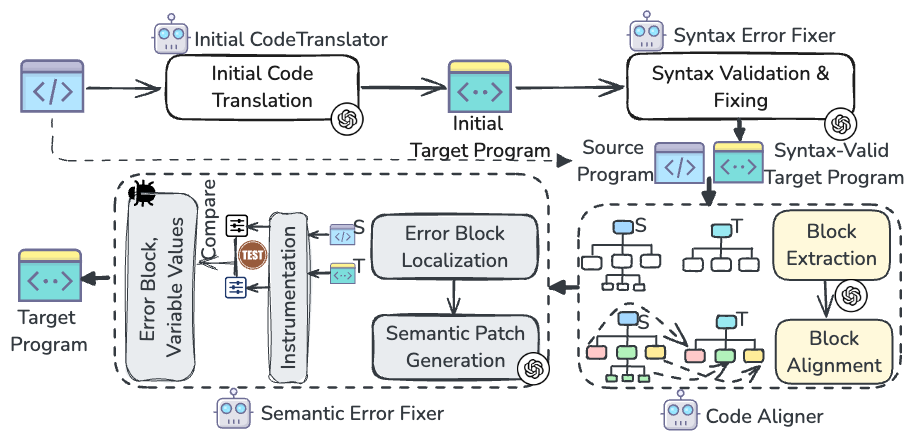}
    \caption{Workflow of \ourtool{}}
    \label{fig:pipline}
\end{figure*}

Figure~\ref{fig:pipline} illustrates the workflow of \ourtool{}, which is a multi-agent system consisting of four different LLM-based agents collaborating with each other, including \textit{Initial Code Translator}, \textit{Syntax Error Fixer}, \textit{Code Aligner}, and \textit{Semantic Error Fixer}. 

\begin{itemize}[leftmargin=10pt, topsep=2pt]
\item \textbf{Initial Code Translator} leverages test cases to enhance LLMs in translating the source program into an initial version of the target program.

\item \textbf{Syntax Error Fixer} aims at iteratively addressing the syntax errors in the target program based on compilation or interpreting error messages with the self-debugging capabilities of LLMs.

\item \textbf{Code Aligner} first divides the source program into blocks based on the control flow, and then leverages LLM to map each block of the source program to that of the target program. The mapping aims at facilitating a fine-grained comparison of runtime behaviors (\eg{} runtime value of specific variables) between the source program and the target program in the following \textit{Semantic Error Fixer} Component. Different from previous code mapping strategies~\cite{transMap} which purely rely on LLMs to perform statement-level alignment, \textit{Code Aligner is novel in incorporating the synergy of both program analysis and LLMs at the block level.}

\item \textbf{Semantic Error Fixer} first localizes the suspicious block in the target program, which exhibits different runtime behaviors from its aligned block in  the source program; and then it leverages LLMs to specifically fix the error block with the observed runtime difference. \textit{Semantic Error Fixer is novel in fixing the semantic errors during code translation in such a fine-grained way.}  
\end{itemize}

In particular, whenever the target program passes all the generated tests, the workflow terminates, and the target program is returned as the final target program; otherwise, the workflow proceeds to fix the syntax or semantic errors of the target program. 




\subsection{Initial Code Translator}
Following the previous code translation work \unitrans{}~\cite{unitrans}, \textit{Initial Code Translator} mainly includes two parts, \ie{} test generation and direct code translation. 
\textit{Test Generation.} As revealed by \unitrans{}~\cite{unitrans}, including test inputs and outputs in the prompt can boost LLM-based code translation. Specifically, we first leverage LLMs to generate test inputs for the given source program with the prompt \textit{``Please generate five inputs for the given source program''}; and the outputs of executing the source program with the generated inputs would be regarded as the test outputs. 
\textit{Direct Code Translation.}  We leverage LLMs to directly generate the target program (\ie{} the initial target program) for the given source program with the generated test inputs and outputs.

\subsection{Syntax Error Fixer}

\textit{Syntax Error Fixer} iteratively leverages LLM to fix the syntax errors in the target program through three steps, \ie{} (i) \textit{syntax validation}, (ii) \textit{fixing strategy planning}, and (iii) \textit{syntax patch generation}. 
\textit{Syntax validation} invokes external tools (\eg{} compilers for Java/C++ or an interpreter for Python) to check the syntactic correctness of the target program. If no syntax errors are found, the target program proceeds to the next agents; otherwise, the process moves to \textit{fixing strategy planning}, where error message is collected and converted into specific fix suggestions. 
As shown in Figure~\ref{fig:syntax_fix}, the fixing suggestion is generated by the LLM based on the collected error message. The underlying inspiration is that planning can enhance the effectiveness of LLM-based agents~\cite{rahardja2025can}. Based on the suggestion, \textit{syntax patch generation} prompts LLMs to create patches to fix the errors.

\textbf{Fixing workflow.} The patched target program would further go to \textit{syntax validation} of the next iteration. The iterative process terminates when (i) there are no syntax errors or (ii) the same syntax errors (via String comparison) occur at the same buggy location as the previous iteration (to prevent an infinite loop). Otherwise, if there are syntax errors different from the previous iteration, \ourtool{} continues the iterative fixing process. 

\begin{figure}[t]
    \centering
    \includegraphics[width=0.95\textwidth]{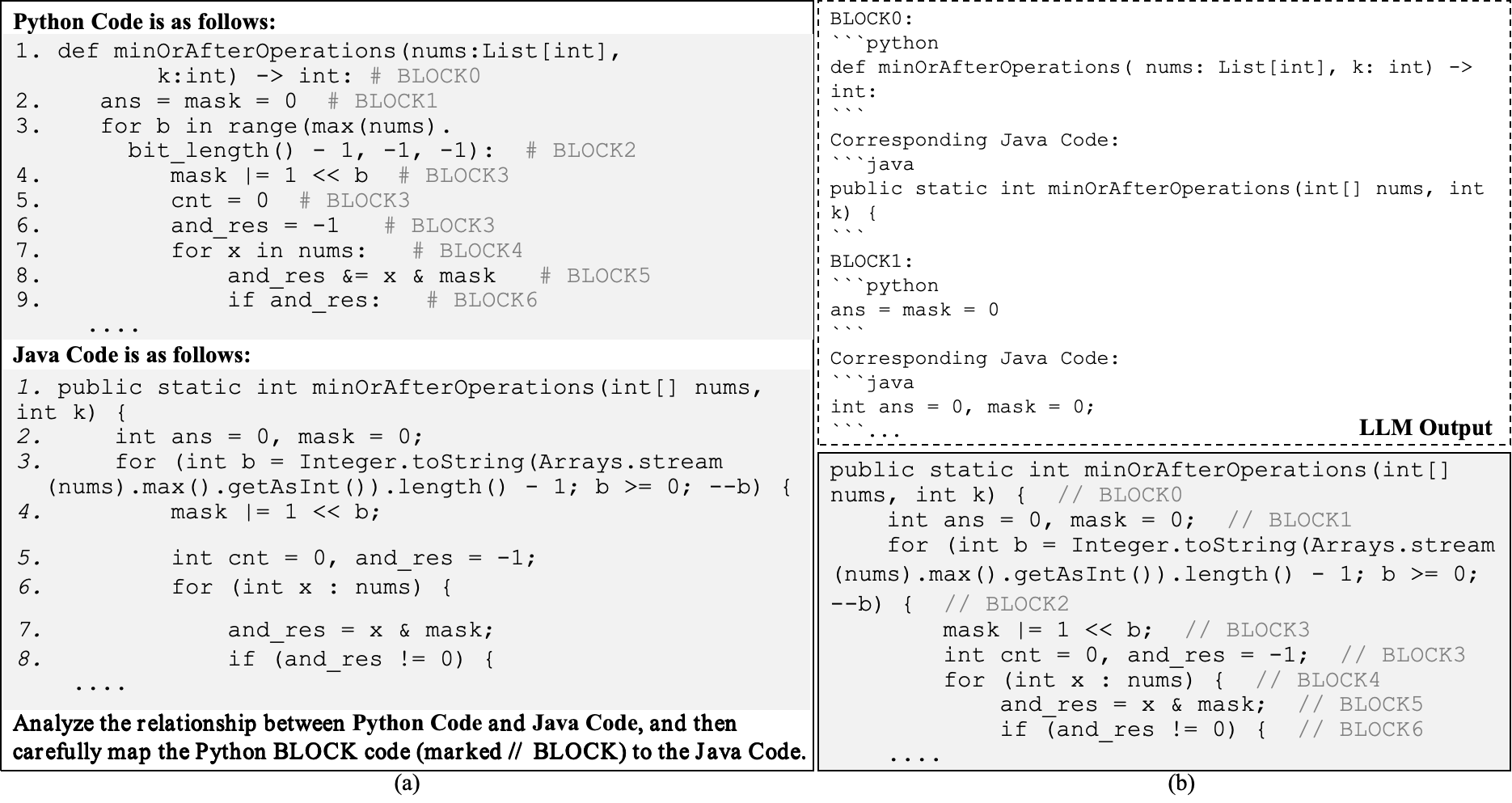}
    \caption{Prompts in \textit{Coder Aligner}}
    \label{fig:codMapping}
\end{figure}

\subsection{Code Aligner}
As shown by the motivating example in Section~\ref{sec:motivate}, directly fixing semantic errors with test inputs and outputs can be challenging for LLMs. Mapping the semantically-equivalent code elements (\ie{} statements or blocks) between the source program and the target program can help localize the error code element, thus narrowing down the fixing space of semantic errors. Therefore, before running \textit{Semantic Error Fixer}, \ourtool{} first includes the LLM-based agent (\ie{} \textit{Code Aligner}) to map semantically-equivalent code elements between the source program and the target program. The previous code mapping technique \transmap{}~\cite{transMap} purely relies on LLMs to perform statement-level mapping, however statement-level mapping can be too fine-grained to be practical, as it is common for (i) one statement aligns (in the source program) with multiple statements (in the target program) or (ii) the order of statements can be very different between the source program and target program. As a result, LLMs exhibit limited mapping accuracy as revealed in the evaluation of \transmap{}~\cite{transMap}.
Therefore, to address these limitations, \textit{Code Aligner} proposes a block-level mapping technique, which aligns code elements in a coarse-grained granularity (\ie{} block-level) with the synergy of both program analysis and LLMs. In particular, \textit{Code Aligner} includes two steps, \ie{} (i) \textit{block extraction} that divides the source program into blocks via control-flow analysis, and (ii) \textit{block alignment} that maps each block in the source program to the target program via LLMs. For better illustration, we denote the source program as $P_S$ and the target program as $P_T$.

\textbf{Block Extraction.} We first construct the control flow graph of the source program and then divide the source program into blocks based on the control flow with the following division criteria.

\begin{itemize}[leftmargin=15pt, topsep=2pt]
\item A continuous sequence of statements that have no jumps in or out of the middle of a block would be regarded as a block. For example, in Figure~\ref{fig:codMapping}.a, Line 4 - 6 is a block (\ie{} marked as BLOCK3). \looseness=-1

\item Any control flow statement (\ie{} the statement that can result in different execution paths, such as \texttt{while}, \texttt{for}, \texttt{try}, or \texttt{if}) would be regarded as a block. For example, in Figure~\ref{fig:codMapping}.a, Line 3 is a block (\ie{} BLOCK2) with a \texttt{for} statement; Line 9 is a block (\ie{} BLOCK6) with an \texttt{if} statement.

\end{itemize}

In fact, the block here is similar to the concept of \textit{basic blocks}~\cite{953283} in the control flow graph. However, the basic block is often on the granularity of a three-address instruction, which can be too fine-grained for the code translation scenario. Therefore, we adjust the scope of the basic block in this work based on the two criteria above. 
We alternatively refer to the blocks in the source program as source blocks and those in the target program as target blocks.
After extraction, the source program is divided into a sequence of numbered blocks, \ie{} $P_S = <B_{S1}, B_{S2}, ... B_{Sn}>$  where $B_{Si}$ denotes the source block in the source program (as shown in Figure~\ref{fig:codMapping}.a).

\textbf{Block Alignment.}
After dividing the source program into blocks, \textit{Code Aligner} further leverages LLMs to map each block to the target program. As shown in Figure~\ref{fig:codMapping}.a, 
LLMs are prompted to map the numbered source blocks to the corresponding target block; the top part of Figure~\ref{fig:codMapping}.b shows the mapping outputs generated by LLMs,  which are further post-processed into a structured representation (as shown in the bottom part of Figure~\ref{fig:codMapping}.b). After the block alignment, the target program is then divided into target blocks, \ie{} $P_T = <B_{T1}, B_{T2}, ... B_{Tn}>$, with mapping function $f_{map}(B_{Si}) = B_{Tj}$.

\begin{figure}[t]
\begin{minipage}{0.7\linewidth}
\scriptsize
\begin{algorithm}[H]\label{algo:localization}
\captionsetup{font=scriptsize}
\caption{ Error Block Localization Algorithm}
\KwIn{$P_S = <B_{S1}, B_{S2}, ... B_{Sn}>$, $P_T = <B_{T1}, B_{T2}, ... B_{Tn}>$, $\mathbb{V}_S=\{V_S^{t_k}\}$, $\mathbb{V}_T=\{V_T^{t_k}\}$, $T=\{t_1, t_2, ..., t_K\}$, $f_{map}$, $f_{id}$}
\KwOut{Error block in target program $B_{Te}$}

\For{$t_k$ in $T$}{
    $l \gets 0$\;
    \While{$l < len(V_T^{t_k})$}{
        $V_S  \gets V_{{S}}^{t_k}[l]$;
        $V_T  \gets V_{{T}}^{t_k}[l]$\;
        $B_{Si} \gets f_{id}(V_S) $;
        $B_{Tj} \gets f_{id}(V_T) $\;
        \If{$V_T == NULL$}{
            \Return $f_{map}(B_{Si})$ \;
        }
        
        \eIf{$B_{Tj} \neq f_{map}(B_{Si})$}{        
            \eIf{$B_{Tj}$ is a control flow statement}{
                \Return $B_{Tj}$ \;
               
            }
            {
                 \Return $f_{map}(B_{Si})$ \;
            }   
        }{
            \eIf{$Equal(V_S, V_T)$}{
                \textbf{continue}\;
            }{
                \Return  $B_{Tj}$ \;
            }
        }
        $l \gets l + 1$\;
    }
    
}

\end{algorithm}
\end{minipage}
\end{figure}


        
               
    

\subsection{Semantic Error Fixer}
Based on the block-level mapping between source and target program, \ourtool{} then performs a fine-grained fixing process by (i) first localizing the error target blocks by comparing the dynamic behaviors of each mapped pair of source blocks and target blocks (\ie{} \textit{Error Block Localization}) and (ii) then specifically fixing the error target block with relevant error information (\ie{} \textit{Semantic Patch Generation}). 
Unlike previous LLM-based code translation works~\cite{unitrans,DBLP:conf/icse/PanIKSWMSPSJ24} that directly leverage LLMs to fix semantic errors without pinpointing the suspicious location, \textit{Semantic Error Fixer} can (i) not only narrow down the fixing space by pinpointing the error target block (ii) but also provide detailed error information about the runtime values within the block rather than only providing the test inputs/outputs of the entire program. We then explain each step in detail.

\subsubsection{Error Block Localization}
For error block localization, \ourtool{} first collects the runtime values of blocks in both source and target programs (\ie{} \textit{Runtime value collection}) and then detects the target block with different values from its mapped source block (\ie{} \textit{Runtime value comparison}).

\textbf{Runtime value collection.} \ourtool{} first collects the runtime values of all the variables within each block for both source and target programs. In particular, \ourtool{} first instruments both source and target programs by adding logging statements at the entry  or exit of each block (more details are in Section~\ref{sec:implement}); then \ourtool{} executes the instrumented source and target program with each test input and collects the runtime values of all the variables within each block for both source and target programs. Specifically, the execution trace of the instrumented source program $P_S$ with test case $t_k$ can be denoted as a list $V_S^{t_k} = <V_{S1}^{t_k}, V_{S2}^{t_k}, ..., V_{SL}^{t_k}>$, where $V_{Sl}^{t_k}$ contains the runtime values within the $l^{th}$ execution instance of the source block $S_{Bi}$, \ie{} $S_{Bi} = f_{id}(V_{Sl}^{t_k})$, and the function $f_{id}$ returns the block of the execution block instance. Additionally, the runtime values of source program $P_S$ with the entire test suite $T$ can be denoted as $\mathbb{V}_S=\{V_S^{t_k}\}$. Similarly, the runtime values of target program $P_T$ with the entire test suite $T$ can be denoted as $\mathbb{V}_T=\{V_T^{t_k}\}$, where $V_T^{t_k} = <V_{T1}^{t_k}, V_{T2}^{t_k}, ..., V_{TL'}^{t_k}>$.

\textbf{Runtime value comparison.} As illustrated in Algorithm~\ref{algo:localization}, \ourtool{} then localizes the error target block by comparing the collected values of each pair of mapped blocks. 
The algorithm iterates the comparison over each test case $t_k$ (Line 1). In particular, along the execution trace of the target program (Line 3), the algorithm compares each  block execution instance iteratively. First, when the runtime values of the current target block do not exist (\ie{} indicating there is some runtime error when the target program executes the block), the algorithm returns the target block that is mapped with the current source block as the error block  (Line 6 - 7). Second, if the current source block and the current target block are not mapped (Line 8), which indicates  there is some mismatching introduced into the control flow, the algorithm returns the control flow statement as the error block. Third, if the current source block and the current target block are mapped and their runtime values are equal (Line 14), which indicates these two blocks are semantically equivalent, the algorithm proceeds to the next iteration; otherwise, the current target block is returned as the error block when the runtime values are not equal between the mapped pair of the source block and the target block.

\subsubsection{Semantic Patch Generation}
After identifying the specific error block within the target program, \ourtool{} leverages LLMs to generate patches for  the error target block. In particular, we include both the vanilla fixing strategy and the value-aware fixing strategy as follows.

\begin{itemize}[leftmargin=10pt, topsep=5pt]
    \item \textit{Vanilla fixing strategy} prompts LLMs to fix the error target block based on static information (\ie{} the code of the error target block and its mapped source block);
    \item \textit{Value-aware fixing strategy} prompts LLMs to fix the error target block by further providing the collected runtime values of the error target block and its mapped source block.

\end{itemize}

These two fixing strategies are complementary, as existing LLMs exhibit imperfect capabilities of reasoning the runtime behaviors of program~\cite{chen2024reasoning}. As a result, runtime values can sometimes be helpful for LLMs to understand bug causes, especially for the cases with extreme values like data overflow (which are the cases value-aware fixing strategies can be helpful for); but sometimes runtime values can be too obscure and overwhelming to negatively limit  LLMs in understanding bugs (which are the cases vanilla fixing strategies can be helpful for). Our ablation study results in Section~\ref{rq:ablation} further confirm the complementarity between these two fixing strategies.

\begin{figure*}[t]
    \centering
    \includegraphics[width=0.95\textwidth]{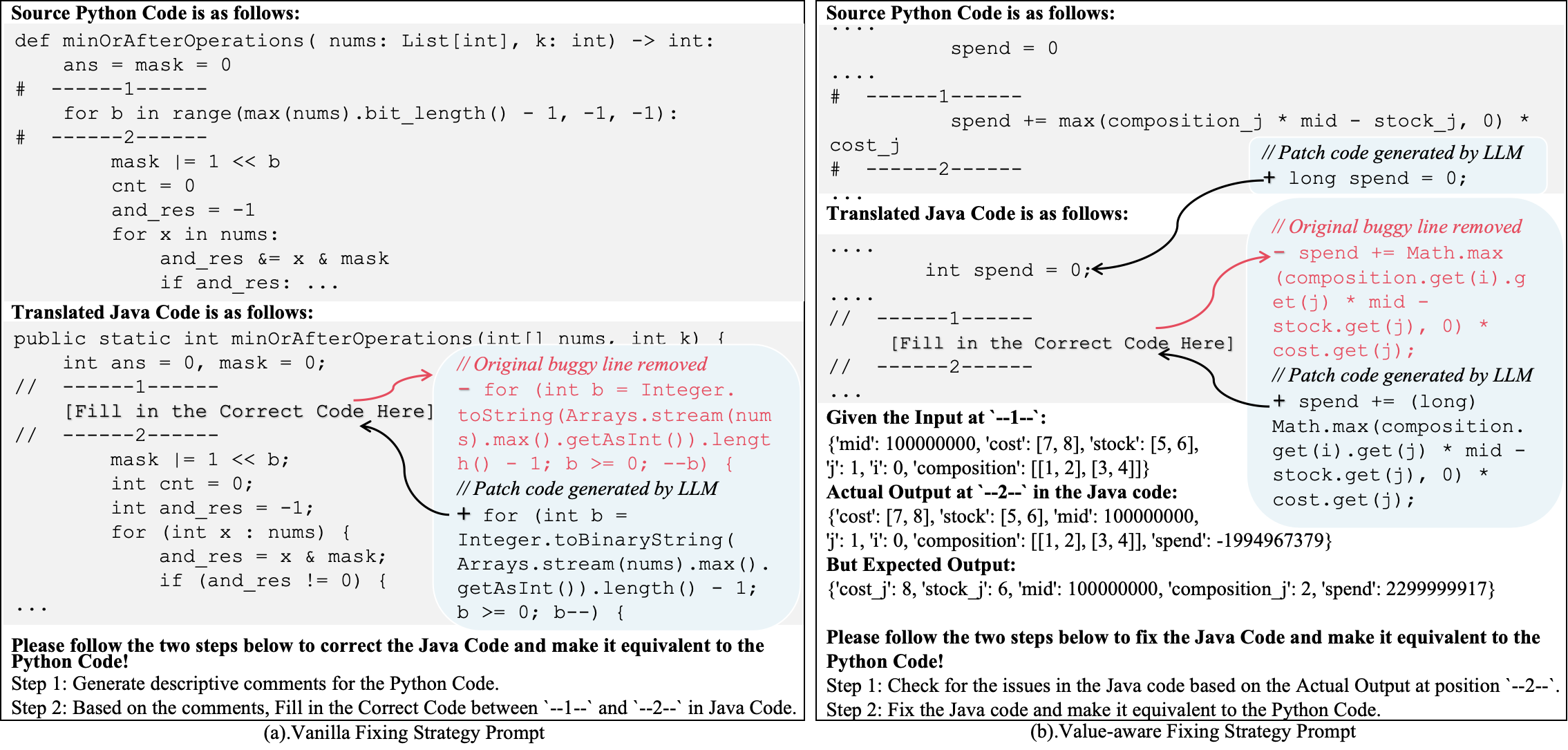}
    \caption{Prompts in Semantic Patch Generation}
    \label{fig:semantic_prompt}
\end{figure*}

For example, Figure~\ref{fig:semantic_prompt}.a and Figure~\ref{fig:semantic_prompt}.b show the prompts used in the vanilla and value-aware fixing strategies, respectively. For both cases, the error block identified in the previous localization step is the code segment between the markers ``--1--'' and ``--2--''. In particular, we adopt a cloze-style fixing prompt by querying LLMs to directly re-generate the correct code (\ie{} ``Fill in the Correct Code Here''), which is commonly used in LLM-based program repair\cite{DBLP:conf/kbse/ZhangFZYSC23,DBLP:conf/sigsoft/0003X023, DBLP:conf/kbse/XiaDZ23, DBLP:conf/sigsoft/XiaZ22}; in addition, both fixing strategies follow Chain-of-Thought reasoning prompts with two steps, which have been shown to be effective in previous research~\cite{DBLP:conf/nips/Wei0SBIXCLZ22, DBLP:conf/iclr/0002WSLCNCZ23, DBLP:journals/corr/abs-2310-04959}.

\textbf{Example illustration.} Figure~\ref{fig:semantic_prompt}.a illustrates the vanilla fixing strategy, which prompts LLMs to generate the correct code for the error target block based on the mapped source block and surrounding contexts.  
Figure~\ref{fig:semantic_prompt}.b illustrates the value-aware fixing strategy. After being translated from Python to Java, the target program encounters a data overflow issue where the \texttt{spend} variable exceeds the range for its type.  The expected output is ``2,299,999,917'' within the mapped source block, but the actual output within the target block is ``-1,994,967,379'' due to the data overflow. By including such extreme runtime values in the prompt, the value-aware fixing strategy can remind LLMs of the potential type errors leading to such extreme runtime values. As a result, the value-aware fixing strategy can fix the error target block by re-declaring \texttt{spend} as the \texttt{Long} type. 

\textbf{Fixing workflow.} During the semantic patch generation, \ourtool{} iteratively applies both fixing strategies. In each iteration, each generated patch would be executed to validate whether the error block of the new target program exhibits no difference in the runtime values compared to the source program. If the runtime value difference of the current error block has been eliminated, \ourtool{} proceeds to fix the next error block (if has any); otherwise, the fixing process terminates by concluding as a failed attempt.

\section{Experimental Setting}

We evaluate \ourtool{} by answering the following research questions:
\begin{itemize}[leftmargin=10pt, topsep=5pt]
\item \textbf{RQ1 (Overall Effectiveness):} How does \ourtool{} compare to state-of-the-art transpilers? 
\item \textbf{RQ2 (Ablation Evaluation):} How does each agent in \ourtool{} boost code translation? 
\item \textbf{RQ3 (Repair Accuracy):} How accurate is \ourtool{} in repairing errors?
\item \textbf{RQ4 (Generalization):} How does \ourtool{} perform with different backbone LLMs?
\end{itemize}

\subsection{Benchmark}


 




\begin{wraptable}{r}{0.5\textwidth}

    \centering
	\small
	\caption{Line Distribution and Average Line Coverage}\label{table:LineDis}
 
\begin{adjustbox}{width=0.5\columnwidth}
\begin{tabular}{c|cc|cc|cc}
\toprule
\multirow{2}{*}{\makecell{Line \\ Distribution}} & \multicolumn{2}{c|}{Python (\%)}           & \multicolumn{2}{c|}{Java (\%)}             & \multicolumn{2}{c}{C++ (\%)}              \\ \cline{2-7}

                                   & \multicolumn{1}{c|}{Pct.} & Cov. & \multicolumn{1}{c|}{Pct.} & Cov. & \multicolumn{1}{c|}{Pct.} & Cov. \\ \midrule
{[}0,5)                            & \multicolumn{1}{c|}{11.0}       & 100.0    & \multicolumn{1}{c|}{2.7}        & 100.0    & \multicolumn{1}{c|}{3.2}        & 100.0    \\ \hline
{[}5,10)                           & \multicolumn{1}{c|}{24.6}       & 100.0    & \multicolumn{1}{c|}{13.1}       & 100.0    & \multicolumn{1}{c|}{13.6}       & 100.0    \\ \hline
{[}10,15)                          & \multicolumn{1}{c|}{35.5}       & 98.2     & \multicolumn{1}{c|}{24.9}       & 99.0     & \multicolumn{1}{c|}{31.2}       & 99.9     \\ \hline
{[}15,20)                          & \multicolumn{1}{c|}{16.2}       & 96.8     & \multicolumn{1}{c|}{26.7}       & 97.6     & \multicolumn{1}{c|}{20.8}       & 100.0    \\ \hline
{[}20,60)                          & \multicolumn{1}{c|}{12.7}       & 95.2     & \multicolumn{1}{c|}{32.6}       & 98.6     & \multicolumn{1}{c|}{31.2}       & 99.0     \\ \bottomrule
\end{tabular}
\end{adjustbox}
\end{wraptable}

\textbf{Limitations of Existing Benchmarks.}
Although there are many existing code translation benchmarks~\cite{DBLP:conf/nips/RoziereLCL20, DBLP:conf/iclr/RoziereZCHSL22, DBLP:conf/nips/LachauxRSL21,DBLP:conf/eacl/AhmadCRC23, DBLP:conf/emnlp/XieNFR23}, they have the following limitations.
First, all existing benchmarks suffer from potential data leakage issues, as their translation tasks are constructed from public programming competitions by the training data timestamp of the most recent LLMs (\eg{} the most widely-used evaluation dataset TransCoder-ST~\cite{DBLP:conf/nips/RoziereLCL20} is created in 2020). Second, some benchmarks involve the translation between only two languages~\cite{DBLP:conf/nips/LuGRHSBCDJTLZSZ21, DBLP:conf/acl/AhmadTCC23, DBLP:conf/sigsoft/NguyenNN13, DBLP:conf/iclr/ChenLS18a}, such as the Java-C\# benchmark CodeTrans~\cite{DBLP:conf/nips/LuGRHSBCDJTLZSZ21} or the Java-Python benchmark AVATAR~\cite{DBLP:conf/acl/AhmadTCC23}, which limits the generalization of evaluation. Third, some benchmarks (\eg{} CodeNet~\cite{DBLP:journals/corr/abs-2105-12655}) suffer from quality issues, and half of its tasks are manually identified as incorrect code by experts ~\cite{DBLP:conf/aaai/Zhu0R22}, thus can harm the soundness of the evaluation.

\textbf{New Benchmark Construction.}
To mitigate the aforementioned limitations, particularly the critical issue of data leakage, we establish a novel benchmark for code translation. Our selection of programming languages for this benchmark focuses on Java, Python, and C++. 
This choice is primarily driven by their long-standing and consistent presence among the top three most popular languages according to the TIOBE index~\cite{tiobe}.
Furthermore, this aligns with the prevailing focus of prior research in code translation~\cite{DBLP:conf/acl/AhmadTCC23, DBLP:conf/nips/LachauxRSL21, DBLP:conf/nips/RoziereLCL20, DBLP:conf/emnlp/XieNFR23,unitrans}, which predominantly concentrates on translation among these three languages. 
Specifically, from the programming competition websites (\eg{} LeetCode~\cite{leetcode} and GeeksforGeeks~\cite{geeksforgeeks}), we collect the solutions of programming tasks for these three programming languages, which are released after \textit{August 2023}. 
As the solutions in these websites typically come with only two or three test cases, which can be insufficient for guaranteeing the semantic correctness of code~\cite{DBLP:conf/nips/LiuXW023}, we further leverage gpt-4o-mini~\cite{gpt4omini} to generate 10 additional test cases per solution, to ensure the sufficiency of tests for each translation task.  We execute the collected code solutions with test cases and discard the tasks whose solutions exhibit inconsistent behaviors between different languages. 
Lastly, two authors of this work further manually check each translation task to ensure the benchmark quality.

\textbf{Benchmark Statistics.}
In this way, we obtain 210 pairs of Python-Java translation tasks, 200 pairs of Python-C++ translation tasks, and 204 pairs of Java-C++ translation tasks.
Table~\ref{table:LineDis} presents the line distribution of our benchmark and its corresponding average line coverage.
Furthermore, the overall average line coverage of each program with test cases achieves 98.4\% for Python, 98.7\% for Java, and 98.4\% for C++, indicating the test sufficiency for each translation task.

\subsection{Baselines}

\textbf{Code Translation Baselines.}  We include the following state-of-the-art LLM-based and learning-based transpilers as baselines.
\begin{itemize}
    \item \textit{\unitrans{}~\cite{unitrans}} is the latest LLM-based code translator, which iteratively fixes translated programs with LLMs. It is notable that another LLM-based technique proposed by Pan et al.~\cite{DBLP:conf/icse/PanIKSWMSPSJ24} also shares a similar fixing approach as \unitrans{}, thus we do not include it as a separate baseline.

    \item \textit{\transcoder{}~\cite{DBLP:conf/nips/RoziereLCL20}} is a representative learning-based code translation technique, which has been evaluated by almost all the previous code translation research~\cite{unitrans, DBLP:conf/iclr/RoziereZCHSL22, DBLP:conf/iclr/SzafraniecRLLCS23, DBLP:conf/eacl/AhmadCRC23, DBLP:conf/nips/LachauxRSL21, DBLP:conf/emnlp/XieNFR23}.
\end{itemize}

\textbf{Program Repair Baseline.} To evaluate the error repair performance of \ourtool{}, we compare it with \textit{\agentless{}}~\cite{Agentless}, a widely evaluated LLM-based program repair approach~\cite{DBLP:journals/corr/abs-2503-21710, DBLP:journals/corr/abs-2409-00899, rahardja2025can}.
We exclude \autocoder{}~\cite{AutoCodeRover}, another commonly evaluated LLM-based method, in our comparison because its coarse-grained, function-level repair strategy is incompatible with our evaluation setting. 
Furthermore, to assess the impact of code mapping on repair effectiveness, we construct a variant named \textit{\ourtool{}\textsubscript{TM}} based on \transmap{}. In this variant, only the \textit{Code Aligner} component of \ourtool{} is replaced with the mapping strategy from \transmap{}, while all other components remain unchanged.

\subsection{Studied Models}
In RQ4, we evaluate \ourtool{} with several open-source LLMs to study its generalization. In particular, we focus on models with fewer than 10 billion parameters and whose training data cutoff predates our benchmark, to avoid data leakage.

\begin{itemize}
\item \textit{\deepseek{}~\cite{deepseek}} with 6.7 billion parameters, initialized from deepseek-Coder-6.7b-base and fine-tuned on 2 billion tokens of instruction data with a knowledge cutoff of February 2023.

\item \textit{\llama{}~\cite{LlamaInstruct}} with 8 billion parameters, which is an instruction-tuned  model from the Llama-3 family, optimized for dialogue usage with a knowledge cutoff of March 2023.

\item \textit{\chatglm{}~\cite{ChatGLM2}} with 6 billion parameters, which is the second version of ChatGLM-6B~\cite{ChatGLM} released in June 2023 and fine-tuned for general-purpose tasks.
\end{itemize}

Additionally, we also include \textit{\DP{}} (33B parameters) in our experiments to evaluate the generalization of \ourtool{} on larger LLMs.

\subsection{Evaluation Metrics}
\textbf{Code Translation Metrics.} Following previous work~\cite{DBLP:conf/iclr/RoziereZCHSL22, DBLP:conf/eacl/AhmadCRC23, DBLP:conf/nips/RoziereLCL20}, we use the following metrics to evaluate the effectiveness of code translation techniques.

\begin{itemize}[topsep=5pt, leftmargin=10pt] 
\item  \textit{Computational Accuracy (CA)}~\cite{DBLP:conf/nips/RoziereLCL20}, the most important metric in code translation, which measures translation accuracy based on functional correctness. CA assesses whether the target program passes all test cases, \ie{} whether the target and source programs produce the same outputs with the same test inputs.

\item \textit{CodeBLEU}~\cite{DBLP:journals/corr/abs-2009-10297}, a metric for the similarity between target and source program.
\end{itemize}

\subsection{Implementation}\label{sec:implement}

\textbf{Baseline Implementation.} 
For \unitrans{}~\cite{unitrans}, we obtain its implementation from its replication package and make the following adjustments for comparison with \ourtool{}. First, we replace the backbone LLM in \unitrans{} with the same LLM used in \ourtool{}. Second, we modify its fixing phase by splitting it into a syntax error fixer and a semantic error fixer, so as to compare with relevant components of \ourtool{}. 
Furthermore,  we align its iteration strategy (\ie{} only iterating within a fixed threshold of iterations) into the same dynamic strategy as \ourtool{}, for fair comparison. For  \transcoder{}~\cite{DBLP:conf/nips/RoziereLCL20} implementation, we directly replicate it with the released implementation with its optimal model weights. We fix the beam\_size parameter at 10 and select the first output to re-evaluate it on our benchmark. 
For \ourtool{}\textsubscript{TM}, we replace the  \textit{Code Aligner} component with the \transmap{}-related implementation. For \agentless{}, we directly use the publicly released implementation.

\textbf{\ourtool{} Implementation}.
For each agent in \ourtool{}, (i) \textit{Initial Code Translator} adopts the same setting as \unitrans{}; (ii) \textit{Syntax Error Fixer} adopts javac for Java, GCC for C++, and the Python interpreter for Python, for syntax validation; (iii) \textit{Code Aligner} adopts Joern~\cite{joern}, a static code analysis tool to generate control flow graphs, which supports multiple languages; (iv) in \textit{Semantic Error Fixer}, for \textit{Error Block Localization}, we insert log statements at either the entry or exit points of each block to capture the runtime values of all variables within the block. If the code block contains a return statement, a log statement is inserted at the entry to capture the return value; otherwise, it is located at the exit. For \textit{Runtime Value Comparison}, we convert the recorded variable values into ``JSON'' format for standardized comparison. Data types such as ``List,'' ``Array,'' and ``Deque'' are mapped to ``JSON'' arrays, and ``int,'' ``float,'' and other numeric types are converted to ``JSON'' numbers.  Such a conversion helps standardize comparison across different programming languages, which might have varying data types and structures. In addition, we also include a one-shot example in all prompts to guide LLMs in generating the required output format.

\textbf{LLM Settings.} For each studied open-source LLM, we use their released model and weights from HuggingFace~\cite{huggingface}. To control randomness in our experiments, we set the parameters to ``temperature=0'' and ``do\_sample=False''.

\subsection{Experimental Procedure}
In this section, we introduce the corresponding evaluation methodology for each research question.

\textbf{RQ1 (Overall Effectiveness).} RQ1 compares the overall code translation effectiveness of \ourtool{} with two baselines (\ie{} \unitrans{}~\cite{unitrans} and \transcoder{}~\cite{DBLP:conf/nips/RoziereLCL20}) in terms of CA and CodeBLEU metrics.

\begin{table*}[t]
    \centering
	\small
	\caption{Translation Effectiveness of Different Transpilers}\label{table:TransEva}
 
    \begin{adjustbox}{width=1.0\columnwidth}
        \begin{tabular}{c|cc|cc|cc|cc|cc|cc}
        \toprule
        \multirow{2}{*}{Transpilers} & \multicolumn{2}{c|}{Java to Python}       & \multicolumn{2}{c|}{Java to C++}          & \multicolumn{2}{c|}{C++ to Java}          & \multicolumn{2}{c|}{C++ to Python}        & \multicolumn{2}{c|}{Python to C++}        & \multicolumn{2}{c}{Python to Java}       \\ \cline{2-13} 
                                     & \multicolumn{1}{c|}{CA(\%)}     & CB & \multicolumn{1}{c|}{CA(\%)}     & CB & \multicolumn{1}{c|}{CA(\%)}     & CB & \multicolumn{1}{c|}{CA(\%)}     & CB & \multicolumn{1}{c|}{CA(\%)}     & CB & \multicolumn{1}{c|}{CA(\%)}     & CB \\ \midrule
        \transcoder{}                   & \multicolumn{1}{c|}{12.1} & 29.3     & \multicolumn{1}{c|}{13.4} & 43.3     & \multicolumn{1}{c|}{41.5} & 47.0     & \multicolumn{1}{c|}{24.5} & 31.1     & \multicolumn{1}{c|}{10.5} & 36.0     & \multicolumn{1}{c|}{2.4}  & 40.1     \\ \hline
        \unitrans{}                     & \multicolumn{1}{c|}{85.0} & 45.3     & \multicolumn{1}{c|}{93.0} & 69.1     & \multicolumn{1}{c|}{65.5} & 77.3     & \multicolumn{1}{c|}{86.0} & 46.5     & \multicolumn{1}{c|}{81.8} & 59.7     & \multicolumn{1}{c|}{56.2} & 65.8     \\ \hline
        \ourtool{}                   & \multicolumn{1}{c|}{\textbf{93.2}} & \textbf{46.0}       & \multicolumn{1}{c|}{\textbf{94.0}} & \textbf{69.2}     & \multicolumn{1}{c|}{\textbf{91.0}} & \textbf{80.5}     & \multicolumn{1}{c|}{\textbf{94.5}} & \textbf{47.1}     & \multicolumn{1}{c|}{\textbf{87.4}} & \textbf{59.9}     & \multicolumn{1}{c|}{\textbf{89.5}} & \textbf{69.8}     \\ \bottomrule
        \end{tabular}
    \end{adjustbox}
\end{table*}

\textbf{RQ2 (Ablation Evaluation).} RQ2 evaluates the contribution of each agent in \ourtool{} (except \textit{Initial Code Translator} as it is a basic component widely used in previous code translation work~\cite{unitrans} but  not the contribution of our approach). For better illustration, we adopt the following abbreviations: \textit{ICT} for \textit{Initial Code Translator}, \textit{SynEF} for \textit{Syntax Error Fixer}, and \textit{SemEF} for \textit{Semantic Error Fixer}. In particular, we investigate the following variants of \ourtool{} for an ablation study. (i) \textit{ICT}: including only \textit{Initial Code Translator} for code translation; (ii) \textit{ICT + SynEF}: including both \textit{ICT} and \textit{SynEF} agents; (iii) \textit{ICT + SynEF + SemEF}: including \textit{ICT}, \textit{SynEF}, and \textit{SemEF} agents. Additionally, we further evaluate the complementarity between two fixing strategies for semantic errors (\ie{} vanilla and value-aware fixing strategies) with the following variants: (iv) \textit{ICT + SynEF + Val}, which applies the value-aware fixing strategy to the \textit{ICT + SynEF} variant, and (v) \textit{ICT + SynEF + Val + Van}, which further applies the vanilla fixing strategy to the \textit{ICT + SynEF + Val} variant.

\textbf{RQ3 (Repair Accuracy).}
RQ3 assesses the program repair performance of \ourtool{} in comparison with \agentless{} and \ourtool{}\textsubscript{TM}. 
Given an initial target program containing translation errors, generated by the \textit{Initial Code Translator}, we apply each approach to repair it.
A repair succeeds if the resulting target program passes all test cases; otherwise, it fails. We use CA as the evaluation metric.

\textbf{RQ4 (Generalization Evaluation).}
RQ4 replaces the default backbone LLM (\ie{} \deepseek{}) in \ourtool{} with three different LLMs (\ie{} \llama, \chatglm{}  and \DP) to evaluate the generalization of \ourtool{}.

\section{EXPERIMENTAL RESULTS}

\subsection{RQ1: Overall Effectiveness}
Table~\ref{table:TransEva} shows the translation performance of \ourtool{} and baselines. Overall, \ourtool{} achieves the best performance across all six translation scenarios, particularly for translations between dynamic and statically typed languages. For example, in the Python-to-Java translation task, \ourtool{} outperforms \transcoder{} in CA by 87.1\% (= 89.5\% - 2.4\%) and \unitrans{} by 33.3\% (= 89.5\% - 56.2\%). 
In addition,  we also conduct the Wilcoxon signed-rank test~\cite{DBLP:reference/stat/ReyN11} to assess the statistical significance of the performance differences. The results yield $p \ll 0.001$ against both \unitrans{} and \transcoder{}, indicating that \ourtool{} significantly outperforms the baselines.

Our manual analysis of the translation results of the \ourtool{} and baselines (\transcoder{} or \unitrans{}) shows that there are no cases in which the baselines (\transcoder{} or \unitrans{}) succeed while \ourtool{} fails.
This result underscores the inherent limitations of learning-based approaches (\ie{} TransCoder), which rely on data-driven learning and are fundamentally constrained by the quality and coverage of their training data. In contrast, LLM-based methods like \ourtool{} and \unitrans{} possess broader generalization capabilities. However, \ourtool{} and \unitrans{} differ in their error correction strategies. \ourtool{} adopts a two-step process, first localizing faults and then performing targeted repairs, while \unitrans{} attempts to fix errors directly without explicit localization, relying solely on its end-to-end capabilities. This lack of precise fault localization ultimately limits the overall translation performance of \unitrans{}.

To better understand the underlying causes of these differences, we conduct a qualitative analysis of cases where the baselines fail while \ourtool{} succeeds.
For \transcoder{}, frequent failures arise from either duplicating source language syntax in the target language or mismatching APIs, two bug categories summarized by~\cite{DBLP:conf/icse/PanIKSWMSPSJ24} that are not observed in \ourtool{}.
For \unitrans{}, the failed translations exhibit two patterns, including modifying code lines unrelated to the root cause of the error and returning the same program as the buggy translated program~\cite{unitrans}.
One potential reason is that, without fine-grained fault localization, \unitrans{} faces challenges in accurately modifying the buggy program based solely on test inputs and outputs. For example, as shown in Figure~\ref{fig:Sem_fix}, \unitrans{} fails to resolve the error because it fails to identify that the fault originates from the third line of the program. In contrast, \ourtool{} accurately detects the fault by analyzing runtime execution to pinpoint the exact error location, and then leveraging an LLM to perform a precise repair.

 
        

\begin{wraptable}{r}{0.5\textwidth}
    \centering
	\small
	\caption{Iteration Proportions in \ourtool{}}\label{table:IterEva}
 
    \begin{adjustbox}{width=0.45\columnwidth}
        \begin{tabular}{c|c|c|c|c}
        \toprule
        \#Iteration & 0 & 1 & 2 & [3,4]  \\ \midrule
        Java to Python           & 89.5\%         & 7.5\%       & 1.5\%       & 1.5\%        \\ \hline
        Java to C++           & 93.3\%        & 1.6\%     & 0.5\%    &  4.7\%     \\ \hline
        C++ to Java           & 89.0\%       & 4.2\%     & 1.6\%     & 5.2\%         \\ \hline
        C++ to Python        & 92.7\%        & 5.2\%      & 2.1\%      & 0.0\%            \\ \hline
        Python to C++        & 88.5\%          & 5.7\%       & 2.1\%      & 3.7\%             \\ \hline
        Python to Java        & 80.9\%        & 6.9\%       & 4.4\%      & 7.8\%            \\ \bottomrule
        
        \end{tabular}
    \end{adjustbox}
\end{wraptable}

\textbf{Cost Evaluation}. We collect the number of iterations and average time costs for \ourtool{}. Table~\ref{table:IterEva} presents the distribution of iteration counts across different translation tasks. In over 92\% of cases, \ourtool{} completes repair in no more than two iterations, indicating low iteration overhead. 
Additionally, a time consumption analysis shows that \ourtool takes an average of 19 seconds per example, while \unitrans{} takes 24 seconds, highlighting the greater efficiency of \ourtool{}.

\begin{mdframed}[linecolor=gray,roundcorner=12pt,backgroundcolor=gray!15,linewidth=3pt,innerleftmargin=2pt, leftmargin=0cm,rightmargin=0cm,topline=false,bottomline=false,rightline = false, skipabove=10pt, skipbelow=10pt]

\textbf{Translation Effectiveness Evaluation Summary:} \ourtool{} is an effective approach for code translation tasks, consistently outperforming existing state-of-the-art transpilers. Meanwhile, \ourtool{} incurs relatively low iteration and time costs.
\end{mdframed}

\subsection{RQ2: Ablation Evaluation}\label{rq:ablation}
Table~\ref{table:componentEva} illustrates the contribution of each agent in \unitrans{} and \ourtool{} to the translation performance.
Overall, the \textit{Syntax Error Fixer} and \textit{Semantic Error Fixer} components in \ourtool{} positively contribute to translation performance. For example, \ourtool{} achieves a 31.0\% improvement over the \textit{Initial Code Translator} in the C++-to-Java task and a 40.0\% improvement in the Python-to-Java scenario.

Both the \textit{Syntax Error Fixer} and the \textit{Semantic Error Fixer} in \ourtool{} yield larger gains in translation performance than their counterparts in \unitrans{}.  For example, in the Python-to-Java task, the \textit{Syntax Error Fixer} in \ourtool{} outperforms that in \unitrans{} by 27.2\% (=32.9\%-5.7\%), while the \textit{Semantic Error Fixer} shows a 6.1\% (=7.1\%-1.0\%) improvement.
The substantial performance gains of the \textit{Syntax Error Fixer} in \ourtool{} over \unitrans{} stem from its planning strategy for handling error messages,  while the advantage of the \textit{Semantic Error Fixer} lies in its fine-grained fault localization prior to repair.

\begin{table*}[t]
    \centering
	\small
	\caption{Performance Comparison of \unitrans{} (\Abbunitrans{}) and \ourtool{} (\Abbourtool{}) Agents
}\label{table:componentEva}
 
    \begin{adjustbox}{width=1.0\columnwidth}
        \begin{tabular}{c|cc|cc|cc|cc|cc|cc}
        \toprule
        \multirow{2}{*}{CA (\%)}          & \multicolumn{2}{c|}{Java to Python}        & \multicolumn{2}{c|}{Java to C++}           & \multicolumn{2}{c|}{C++ to Java}           & \multicolumn{2}{c|}{C++ to Python}         & \multicolumn{2}{c|}{Python to C++}         & \multicolumn{2}{c}{Python to Java}        \\ \cline{2-13} 
                                     & \multicolumn{1}{c|}{\Abbunitrans{}} & \Abbourtool{} & \multicolumn{1}{c|}{\Abbunitrans{}} & \Abbourtool{} & \multicolumn{1}{c|}{\Abbunitrans{}} & \Abbourtool{} & \multicolumn{1}{c|}{\Abbunitrans{}} & \Abbourtool{} & \multicolumn{1}{c|}{\Abbunitrans{}} & \Abbourtool{} & \multicolumn{1}{c|}{\Abbunitrans{}} & \Abbourtool{} \\ \midrule
        ICT                  & \multicolumn{2}{c|}{84.5}                & \multicolumn{2}{c|}{89.1}                & \multicolumn{2}{c|}{60.0}                & \multicolumn{2}{c|}{83.0}                & \multicolumn{2}{c|}{76.8}                & \multicolumn{2}{c}{49.5}                \\ \hline
        ICT+SynEF          & \multicolumn{1}{c|}{\textbf{85.0}}   & \textbf{85.0}     & \multicolumn{1}{c|}{\textbf{91.0}}   & \textbf{91.0}     & \multicolumn{1}{c|}{64.8}   & \textbf{87.9}     & \multicolumn{1}{c|}{86.0}   & \textbf{87.5}     & \multicolumn{1}{c|}{79.0}   & \textbf{81.3}     & \multicolumn{1}{c|}{55.2}   & \textbf{82.4}     \\ \hline
        ICT+SynEF+SemEF        & \multicolumn{1}{c|}{85.0}   & \textbf{93.2}     & \multicolumn{1}{c|}{93.0}   & \textbf{94.0 }    & \multicolumn{1}{c|}{65.8}   & \textbf{91.0}     & \multicolumn{1}{c|}{86.0}   & \textbf{94.5}     & \multicolumn{1}{c|}{81.0}   & \textbf{87.4}     & \multicolumn{1}{c|}{56.2}   & \textbf{89.5}     \\ \hline

        $\Delta_{\text{ICT+SynEF/Val}}$\%    & \multicolumn{1}{c|}{-}        & 6.3     & \multicolumn{1}{c|}{-}        & 2.5     & \multicolumn{1}{c|}{-}        & 2.1     & \multicolumn{1}{c|}{-}        & 6.5     & \multicolumn{1}{c|}{-}        & 4.6     & \multicolumn{1}{c|}{-}        & 4.3     \\ \hline

        $\Delta_{\text{ICT+SynEF+Val/Van}}$\%      & \multicolumn{1}{c|}{-}        & 1.9     & \multicolumn{1}{c|}{-}        & 0.5     & \multicolumn{1}{c|}{-}        & 1.0     & \multicolumn{1}{c|}{-}        & 0.5     & \multicolumn{1}{c|}{-}        & 1.5     & \multicolumn{1}{c|}{-}        & 2.8     \\ \bottomrule

        \end{tabular}
    \end{adjustbox}

\end{table*}

The \textit{Semantic Error Fixer} in \ourtool{} is more effective at repairing translation errors between dynamic and static languages than between dynamic languages. For example, in the Java-to-C++ scenario, the \textit{Semantic Error Fixer} improves the performance of \ourtool{} by 3.0\%, whereas in the Python-to-Java scenario it delivers a 7.1\% improvement.
This difference arises from the structural gaps between dynamic and static languages. Dynamic languages such as Python often employ compact constructs, for example, ``stnum = "".join(sorted(str(num)))'', which combine multiple operations (\eg{} \textit{type conversion}, \textit{string manipulation}, \textit{sorting}, and \textit{aggregation}) into a single line. 
When performing program translation, LLMs struggle to disentangle the densely packed operations in such expressions, often resulting in semantic mismatches.
In this case, the \texttt{sorted} function in Python defaults to ascending order, but the \textit{ICT} incorrectly translates it into \texttt{Arrays.sort(.., Collections.reverseOrder())}, which produces descending order and breaks semantic consistency.
To address such errors, the \textit{Semantic Error Fixer} compares runtime states to identify inconsistencies. During repair, it then provides test cases (\eg{} input \texttt{num = 2134} with expected output \texttt{1234}) that reveal the correct behavior, enabling the LLM to generate accurate fixes.

Both Value-aware and Vanilla fixing strategies in \textit{Semantic Error Fixer} can improve the overall performance of \ourtool{} in code translation. For instance, in the Java-to-Python translation scenario, the Value-aware strategy increases CA by 6.3\% over \textit{ICT + SynEF}, while the Vanilla strategy provides an additional 1.9\% improvement. This shows that the two fixing strategies are complementary, enhancing the ability of \ourtool{} to correct errors and improve translation performance.

\begin{mdframed}[linecolor=gray,roundcorner=12pt,backgroundcolor=gray!15,linewidth=3pt,innerleftmargin=2pt, leftmargin=0cm,rightmargin=0cm,topline=false,bottomline=false,rightline = false, skipabove=10pt, skipbelow=10pt]
\textbf{Ablation Evaluation Summary:} 
Both the \textit{Syntax Error Fixer} and the \textit{Semantic Error Fixer} in \ourtool{} contribute positively to improving the translation performance of LLMs. Moreover, these components are more effective than their counterparts in \unitrans{}.
\end{mdframed}

 

\begin{table*}[t]
    \centering
	\small
	\caption{Repair Performance Comparison of  \agentless{}, \ourtool{}\textsubscript{TM} and \ourtool{}}\label{table:map}
 
    \begin{adjustbox}{width=1.0\columnwidth}
        \begin{tabular}{c|c|c|c|c|c|c}
        \toprule
        CA (\%)  & Java to Python & Java to C++ & C++ to Java & C++ to Python & Python to C++ & Python to Java \\ \midrule
        \agentless{}      & 9.3            & 18.2        & 2.5          & 20.6         & 2.2           & 0.9            \\ \hline
        \ourtool{}\textsubscript{TM} & 37.5           & 31.8        & 77.2          & 55.9        & 40.9          & 63.2           \\ \hline
        \ourtool{}     & \textbf{59.4}           & \textbf{45.4}        & \textbf{83.5}          & \textbf{76.5}        & \textbf{50.0}          & \textbf{79.2}           \\ \bottomrule
        \end{tabular}
    \end{adjustbox}
\end{table*}
\subsection{RQ3: Repair Accuracy}
Table~\ref{table:map} presents the program repair performance of \ourtool{} and the baselines. The results show that \ourtool{} consistently outperforms both \agentless{} and \ourtool{}\textsubscript{TM}. Overall, \ourtool{} achieves an average repair accuracy that is 56.7\% higher than \agentless{} and 14.5\% higher than \ourtool{}\textsubscript{TM}, which employs the \transmap{} mapping strategy. 
The Wilcoxon Signed-Rank Test~\cite{DBLP:reference/stat/ReyN11} yields
\textit{$p = 1.5 \times 10^{-58}$} versus \agentless{} and
\textit{$p = 1.0 \times 10^{-4}$} versus \ourtool{}\textsubscript{TM},
indicating that \ourtool{} significantly outperforms the baselines in error repair ($p \ll 0.001$).
This advantage stems from two key innovations: a hybrid code mapping strategy that integrates program analysis with LLMs for source-to-target alignment, and a fault localization method that leverages intermediate runtime states to identify and fix errors.

\begin{figure}[t]
    \centering
    \includegraphics[width=0.55\textwidth]{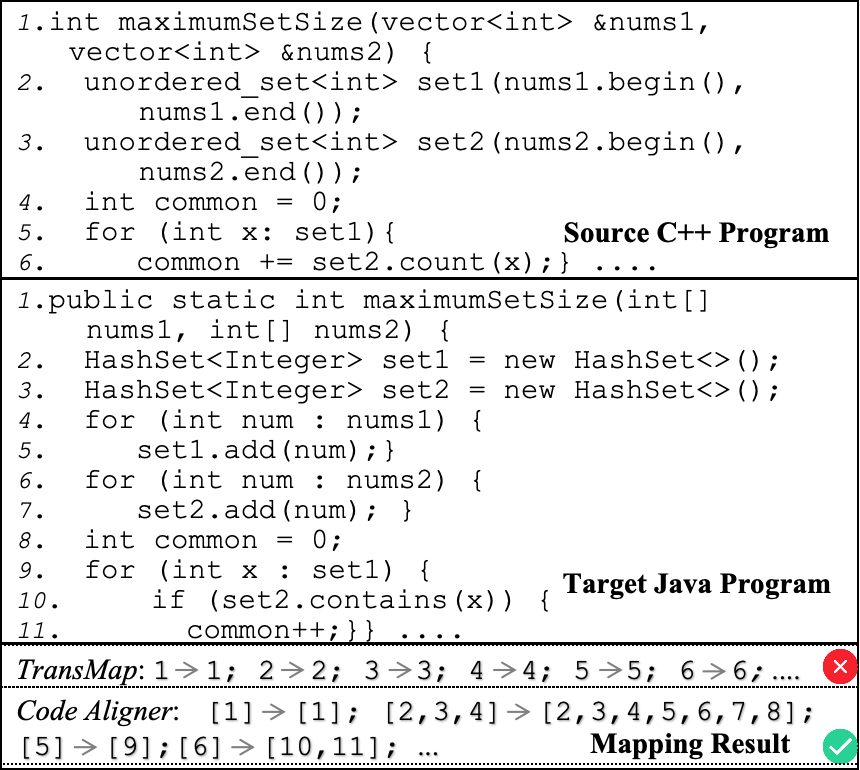}
     \caption{Mapping Example: \transmap{} vs. \textit{Code Aligner}}
    \label{fig:MappingExample}
\end{figure}

For code mapping, the \textit{Code Aligner} in \ourtool{} employs a hybrid strategy that combines CFG analysis with LLM-based alignment. It first decomposes the source program into semantically coherent code blocks and then aligns them using the LLM, effectively handling complex scenarios such as line shifts. For example, as shown in Figure~\ref{fig:MappingExample}, the \textit{Code Aligner} groups C++ lines \texttt{2–4} as a logical block and accurately maps them to Java lines \texttt{2–8}, demonstrating robustness to non-sequential and interleaved mappings. In contrast, \transmap{} in \ourtool{}\textsubscript{TM} relies solely on sequential LLM-based alignment, which is prone to mapping errors in Figure~\ref{fig:MappingExample}, for instance, by incorrectly aligning line \texttt{5} in the source with line \texttt{5} in the target, thereby reducing the effectiveness of fault localization and repair.

For fault localization, \ourtool{} analyzes the intermediate runtime states of both the source and target programs, enabling it to precisely identify the root causes of errors rather than relying on coarse-grained error signals. In contrast, \agentless{} depends solely on end-to-end LLM reasoning for both localization and repair. When only high-level error information is available, \agentless{} must infer the root cause directly, which is challenging for current LLMs and results in substantially lower repair accuracy.

\begin{mdframed}[linecolor=gray,roundcorner=12pt,backgroundcolor=gray!15,linewidth=3pt,innerleftmargin=2pt, leftmargin=0cm,rightmargin=0cm,topline=false,bottomline=false,rightline = false, skipabove=10pt, skipbelow=10pt]
\textbf{Repair Accuracy Evaluation Summary:} 
\ourtool{} achieves substantial program repair performance by combining hybrid code mapping with runtime-guided fault localization, effectively overcoming key limitations of existing LLM-based repair methods.
\end{mdframed}

\subsection{RQ4: Generalization Evaluation}

\begin{table*}[t]
    \centering
	\small
	\caption{Generalization of \ourtool{} in \llama{} (L3), \chatglm{} (CG), and \DP{} (D33)}\label{table:diffLLM}
 
    \begin{adjustbox}{width=1.0\columnwidth}
\begin{tabular}{c|ccc|ccc|ccc|ccc|ccc|ccc}
\toprule
\multirow{2}{*}{CA (\%)}                                   & \multicolumn{3}{c|}{Java to Python}                                & \multicolumn{3}{c|}{Java to C++}                                   & \multicolumn{3}{c|}{C++ to Java}                                   & \multicolumn{3}{c|}{C++ to Python}                                 & \multicolumn{3}{c|}{Python to C++}                                 & \multicolumn{3}{c}{Python to Java}                              \\ \cline{2-19} 
                                                           & \multicolumn{1}{c|}{L3}     & \multicolumn{1}{c|}{CG}     & D33    & \multicolumn{1}{c|}{L3}     & \multicolumn{1}{c|}{CG}     & D33    & \multicolumn{1}{c|}{L3}     & \multicolumn{1}{c|}{CG}     & D33    & \multicolumn{1}{c|}{L3}     & \multicolumn{1}{c|}{CG}     & D33    & \multicolumn{1}{c|}{L3}     & \multicolumn{1}{c|}{CG}     & D33    & \multicolumn{1}{c|}{L3}     & \multicolumn{1}{c|}{CG}     & D33  \\ \midrule
ICT                                                        & \multicolumn{1}{c|}{72.9} & \multicolumn{1}{c|}{18.4} & 93.2 & \multicolumn{1}{c|}{75.1} & \multicolumn{1}{c|}{7.5}  & 89.3 & \multicolumn{1}{c|}{53.5} & \multicolumn{1}{c|}{6.0}  & 78.0 & \multicolumn{1}{c|}{79.5} & \multicolumn{1}{c|}{16.5} & 94.0 & \multicolumn{1}{c|}{66.5} & \multicolumn{1}{c|}{4.5}  & 89.0 & \multicolumn{1}{c|}{29.7} & \multicolumn{1}{c|}{5.4}  & 72.2 \\ \hline
ICT+SynEF                                                  & \multicolumn{1}{c|}{78.7} & \multicolumn{1}{c|}{19.3} & 93.7 & \multicolumn{1}{c|}{81.1} & \multicolumn{1}{c|}{14.4} & 95.4 & \multicolumn{1}{c|}{70.5} & \multicolumn{1}{c|}{7.0}  & 95.0 & \multicolumn{1}{c|}{83.5} & \multicolumn{1}{c|}{17.0} & 95.5 & \multicolumn{1}{c|}{75.5} & \multicolumn{1}{c|}{6.0}  & 92.0 & \multicolumn{1}{c|}{55.0} & \multicolumn{1}{c|}{7.2}  & 86.1 \\ \hline
\begin{tabular}[c]{@{}c@{}}ICT+SynEF+SemEF\end{tabular} & \multicolumn{1}{c|}{87.9} & \multicolumn{1}{c|}{48.8} & 96.6 & \multicolumn{1}{c|}{86.6} & \multicolumn{1}{c|}{20.4} & 95.9 & \multicolumn{1}{c|}{77.0} & \multicolumn{1}{c|}{11.5} & 96.0 & \multicolumn{1}{c|}{90.5} & \multicolumn{1}{c|}{55.0} & 98.5 & \multicolumn{1}{c|}{78.5} & \multicolumn{1}{c|}{13.5} & 93.5 & \multicolumn{1}{c|}{61.2} & \multicolumn{1}{c|}{14.9} & 91.9 \\ \bottomrule
\end{tabular}

    \end{adjustbox}
\end{table*}

Table~\ref{table:diffLLM} presents the translation performance of each component in \ourtool{} across different LLMs. 
The results indicate that \ourtool{} generalizes effectively, consistently enhancing the code translation performance of all LLMs. Even for the large-scale \DP{}, \ourtool{} provides substantial gains. For example, in the Python-to-Java task, \ourtool{} improves performance by 31.5\% (=61.2\%-29.7\%) with \llama{}, by 9.5\% (=14.9\%-5.4\%) with \chatglm{}, and by 19.7\% (=91.9\%-72.2\%) with \DP{}.

Both the \textit{Syntax Error Fixer} and the \textit{Semantic Error Fixer} demonstrate strong adaptability across different LLMs, contributing to consistent improvements in translation performance. The \textit{Syntax Error Fixer} achieves gains of 25.3\% with \llama{} in Python-to-Java and 6.9\% with \chatglm{} in Java-to-C++. Similarly, the \textit{Semantic Error Fixer} yields improvements of 9.2\% with \llama{} in Java-to-Python and 38.0\% with \chatglm{} in C++-to-Python. These results indicate that each component of \ourtool{} generalizes effectively across models, thereby enhancing overall translation performance.

\begin{mdframed}[linecolor=gray,roundcorner=12pt,backgroundcolor=gray!15,linewidth=3pt,innerleftmargin=2pt, leftmargin=0cm,rightmargin=0cm,topline=false,bottomline=false,rightline = false, skipabove=10pt, skipbelow=10pt]
\textbf{Generalization Evaluation Summary:} \ourtool{} can be applied to different LLMs to enhance their performance in code translation tasks.
\end{mdframed}


\section{Discussion}

\subsection{Translation Performance Impact Analysis}

\textbf{Code Complexity Impact.}
We analyze how source code complexity, measured by lines of code (LOC) and cyclomatic complexity (CC), affects the translation performance of \ourtool{} and baselines, as shown in Table~\ref{table:dic_complexity}.
\textit{First, the translation performance of \ourtool{} remains stable as LOC or CC increases.} 
For instance, in the LOC group from [31, 45] to [46, 60], \ourtool{}’s CA increases from 84.6\% to 100.0\%, and in the CC group from [11, 15] to [16, 20], CA increases from 89.5\% to 100.0\%.
\textit{Second, for the same ranges of LOC or CC, \ourtool{} consistently outperforms the baseline.} For example, in the LOC range [31, 45], \ourtool{} achieves a CA that is 78.4\% higher than \transcoder{} and 7.7\% higher than \unitrans{}. This performance advantage primarily results from \ourtool{} performing fine-grained fault localization by comparing intermediate runtime states between the source and target programs, which enables precise correction of translation errors.
In contrast, \unitrans{} and \transcoder{} rely solely on end-to-end semantic reasoning and cannot exploit the intermediate states of the target program. Consequently, they fail to address subtle discrepancies that emerge only at runtime, such as precision-related issues~\cite{unitrans}. As illustrated in Figure~\ref{fig:semantic_prompt}.a, the translated program incorrectly infers the type of the variable \texttt{spend} as \texttt{int} instead of the expected \texttt{long}.
Neither \unitrans{} nor \transcoder{} can correct this error. In contrast, by analyzing and comparing the intermediate states of the source and target programs, \ourtool{} detects the divergence in the runtime values of \texttt{spend},  pinpoints the faulty statement, and leverages the mismatch between the observed and expected outputs to perform precise correction.

\begin{table*}[t]

\caption{Translation Performance (CA\%) of \ourtool{} and Baselines across Line of Code (LOC) and Cyclomatic Complexity (CC)}\label{table:dic_complexity}
\begin{adjustbox}{width=0.9\columnwidth}

\begin{tabular}{cccc|cccc}
\toprule
LOC Groups  & \transcoder{} & \unitrans{} & \ourtool{} & CC Groups   & \transcoder{} & \unitrans{} & \ourtool{} \\ \midrule
(1,15{]}    & 19.8       & 75.6     & 91.9       & (1,5{]}     & 20.3       & 75.8     & 92.1       \\
{[}16,30{]} & 16.0       & 81.1     & 92.2       & {[}6,10{]}  & 13.4       & 81.3     & 90.8       \\
{[}31,45{]} & 6.2        & 76.9     & 84.6       & {[}11,15{]} & 4.2        & 79.2     & 89.5       \\
{[}46, 60)  & 0.0        & 84.6     & 100.0      & {[}16, 20)  & 0.0        & 83.3     & 100.0      \\ \bottomrule
\end{tabular}

 \end{adjustbox}
\end{table*}

\ourtool{} demonstrates stable performance on complex function-level code and thus shows promising potential for project-level extension. Existing project-level translation approaches generally adopt a decompositional paradigm, in which a project is first decomposed into smaller units (typically functions) and then translated in a bottom-up order following the call graph~\cite{DBLP:journals/pacmse/IbrahimzadaKPAPSJ25, DBLP:journals/corr/abs-2503-17741, guan2025repotransagentmultiagentllmframework}. 
Under this paradigm, \ourtool{} can be naturally integrated into project-level workflows by being applied to individual functions during translation, thereby providing fine-grained runtime feedback and repair at the unit level. By improving per-function correctness and stability, \ourtool{} enhances the overall reliability of project-level translation.

\begin{table*}[t]

\caption{Translation Performance on Existing Benchmark}\label{table:dic_evaluBench}
\begin{adjustbox}{width=0.9\columnwidth}

\begin{tabular}{c|c|c|c|c|c|c}
\toprule
CA(\%)     & Java to Python & Java to C++ & C++ to Java & C++ to Python & Python to C++ & Python to Java \\ \midrule
\transcoder{} & 51.0        & 85.5     & 71.2     & 46.0       & 63.6       & 37.8        \\ \hline
\sda{}  & 58.1        & 84.7     & 66.3     & 55.4       & 41.2       & 42.6        \\ \hline
\unitrans{}   & 92.5        & 65.8     & 94.0     & 92.3       & 59.0       & 89.8        \\ \hline
\ourtool{} & 95.7        & 98.0     & 95.2     & 94.7       & 96.2       & 93.3        \\ \bottomrule
\end{tabular}
 \end{adjustbox}
\end{table*}

 

\textbf{Cross-Language API Mapping Impact.}
\ourtool{} performs cross-language core API mapping by leveraging LLMs’ pre-trained knowledge of standard library APIs across different programming languages. Since LLMs are pretrained on large-scale multilingual code and documentation corpora, they exhibit strong capabilities in understanding common APIs~\cite{DBLP:conf/icse/WangLPLL25, DBLP:conf/wcre/ChenGZ0WX24, DBLP:journals/corr/abs-2409-15228} (\eg{} Python’s \texttt{math} and \texttt{collections}, Java’s \texttt{java.io.*} and \texttt{java.lang.*}, and C++’s \texttt{<iostream>} and \texttt{<queue>}) within given code contexts. For instance, in our benchmark, \ourtool{} successfully identifies the \texttt{count()} function in the Python expression \texttt{(nums.count(m)+1)//2} and correctly maps it to Java’s \texttt{java.util.Collections.frequency()} as \texttt{(Collections.frequency(nums,m)+1)/2}.
In addition, during the error-fixing stage, \ourtool{} guides LLMs to correct mistranslations in API usage.
For example, in the statement ``stnum = "".join(sorted(str(num)))'', the densely packed operations cause LLMs to incorrectly translate \texttt{sorted} function as \texttt{Arrays.sort(.., Collections.reverseOrder())}.
In the error-fixing stage, supplying \texttt{num=2134} with the expected output \texttt{stnum=1234} guides LLMs to produce the correct translation, \texttt{Arrays.sort(numChars)}.

    %

\begin{table*}[htb]
    \centering
	\small
	\caption{Comparison of Block Mapping Accuracy Between \ourtool{} and \transmap{}}\label{table:block_map}
 
    \begin{adjustbox}{width=0.9\columnwidth}
    
\begin{tabular}{c|c|c|c|c|c|c}
\toprule
Mapping Accuracy(\%) & Java to Python & Java to C++ &  C++ to Java & C++ to Python & Python to C++ & Python to Java \\ \midrule
\transmap{}             & 64.6        & 79.2            & 58.3     & 80.2 & 93.8       & 67.7        \\ \hline
\ourtool{}           & 95.8        & 97.9            & 97.9    & 99.0 & 100.0      & 96.9        \\ \bottomrule
\end{tabular}

\end{adjustbox}
\end{table*}

\textbf{Block Mapping Impact.}
Block-level mapping plays a critical role in improving translation correctness by providing fine-grained structural alignment between source and target programs. 
To assess the reliability of the mapping results, we compare \ourtool{} with \transmap{} via manual inspection. In particular, we manually inspect 290 mapped source--target block pairs (95\% confidence level, 0.05 margin of error) and report the mapping accuracy in Table~\ref{table:block_map}. Overall, \ourtool{} achieves 95.8\%--100\% mapping accuracy across all six language pairs, substantially outperforming \transmap{} (64.6\%--93.8\%). This accuracy gap indicates that \ourtool{} provides more reliable alignment signals, which are crucial for guiding the subsequent error fixing.
Moreover, occasional mapping failures are unlikely to mislead the subsequent repair process. This is because \ourtool{} compares only the runtime values of variables with the same names within aligned blocks, thereby naturally limiting the impact of imperfect alignments. For example, between Python’s ``\texttt{a,b=0,1}'' and Java’s ``\texttt{a=0}'', only the shared variable ``\texttt{a}'' is used for runtime comparison, preventing unintended mismatches.

In addition, we compare \ourtool{} with \batfix{}~\cite{DBLP:journals/tosem/RamosLMMG24}, which matches source and target control-flow graphs via a MaxSAT-based formulation. Since \batfix{} only supports Java-to-C++ and Python-to-C++, we conduct comparisons on these two language pairs. Overall, \ourtool{} achieves substantially higher mapping accuracy than \batfix{}, reaching 97.0\% vs.\ 73.0\% on Java-to-C++ and 98.0\% vs.\ 32.0\% on Python-to-C++.

\subsection{Existing Benchmark Evaluation}

In the experiments, we collect a new dataset to assess the translation performance of \ourtool{} and the baselines, mitigating potential data leakage issues. 
To further evaluate the generalization of \ourtool{} across different datasets, we use the benchmark released by Roziere et al.\cite{DBLP:conf/nips/RoziereLCL20}, which is widely evaluated in prior work~\cite{DBLP:conf/nips/RoziereLCL20, DBLP:conf/iclr/RoziereZCHSL22, DBLP:conf/eacl/AhmadCRC23, DBLP:conf/oopsla/KaraivanovRV14,DBLP:conf/iclr/ChenLS18a, unitrans}.
Additionally, we include a new baseline, \sda{}~\cite{DBLP:conf/icse/LiuLZ23}, which is a learning-based method augmented with symbolic analysis.
Table~\ref{table:dic_evaluBench} presents the translation performance of \transcoder{}, \sda{} (using the results reported in the original paper, as the reproduction packages are unavailable~\cite{DBLP:conf/icse/LiuLZ23}), \unitrans{}, and \ourtool{} on this benchmark. 
As shown in the table, \ourtool{} consistently outperforms all baselines on this benchmark, highlighting its robustness and generalization ability. 
Furthermore, \ourtool{} achieves a CA of up to 98.0\% on this dataset, translating nearly all programs correctly. This highlights the necessity of constructing a new evaluation dataset to reliably assess translation performance.

\section{Threats to Validity}
We categorize threats to the effectiveness of this work into internal effectiveness and external effectiveness.
(i) One threat to validity is potential bugs in the code implementation of \ourtool{}, which could lead to translation failures. To mitigate this, we use instances of translation failures to debug and improve the implementation.
(ii) Another threat is potential data leakage due to overlap between the evaluation dataset and the training data of LLMs. To address this, we manually construct a new dataset from a time frame after the knowledge cutoff date of LLM, specifically after \textit{August 2023}. We also design comprehensive test cases and calculate line coverage to ensure the equivalence of the constructed source and target code.
(iii) Another threat to internal effectiveness is that \ourtool{} assumes the source program is executable, as it relies on runtime feedback to perform block-level alignment and to validate repairs. If the source code is not runnable (\eg{} missing dependencies, incomplete environments, or uncompilable projects), \ourtool{} may fail to obtain reliable execution signals and thus cannot proceed. This executability assumption is also common in execution-guided translation/repair approaches~\cite{unitrans, DBLP:conf/icse/PanIKSWMSPSJ24, DBLP:journals/corr/abs-2412-14234, transMap}. To mitigate this threat in our evaluation, we only include instances whose source programs can be executed in a controlled environment. Regarding test data, \ourtool{} does not rely on pre-existing test suites, as it can automatically generate additional tests when needed.
(iv) Another threat involves the reproduction of baseline and the calculation of metrics. To minimize these threats, we strictly follow the reproduction documentation and use the source code provided for the baseline. All baselines are evaluated consistently on our collected dataset, and we reuse the code provided in~\cite{DBLP:conf/eacl/AhmadCRC23} for implementing evaluation metrics like CodeBLEU.

\section{Related Work}~\label{sec:related}

\subsection{Code Translation}
Early code translation research relied on program analysis techniques with manually formulated rules, such as C2Rust~\cite{c2rust} for translating C into Rust, and Sharpen~\cite{sharpen} and JavaSharp~\cite{JavaToCSharp} for converting Java to C\#. However, these methods are time-intensive and limit the accuracy and readability of the translated programs~\cite{DBLP:conf/nips/RoziereLCL20}. Learning-based approaches have since emerged~\cite{DBLP:conf/oopsla/KaraivanovRV14, DBLP:conf/kbse/OdaFNHSTN15}, significantly improving over rule-based methods but still struggling with data scarcity and the time-consuming training process~\cite{DBLP:conf/iclr/SzafraniecRLLCS23, DBLP:conf/aaai/Zhu0R22, unitrans}.

LLMs have shown promise in software engineering tasks~\cite{abs240902977}, such as code generation~\cite{DBLP:journals/tosem/HuangYXPXL24,yang2025bamas,LiuCLZHH0D0025}, program repair~\cite{DBLP:conf/icse/FanGMRT23, DBLP:conf/icse/XiaWZ23,guo2026eet,rahardja2025can}, and code summarization~\cite{DBLP:conf/kbse/AhmedD22, DBLP:journals/corr/abs-2304-11384}.
Pan et al.~\cite{DBLP:conf/icse/PanIKSWMSPSJ24} identify key syntactic and semantic challenges in LLM-based code translation. Therefore, adopting the ``translate-then-fix'' paradigm is an effective way to enhance LLM-based translation performance. 
\unitrans{}~\cite{unitrans}, a \sota{} method, adopts this paradigm to conduct code translation. However, the error-fixing stage suffers from poor performance due to the lack of detailed runtime error information.
\transmap{} improves the manual fixing process by aligning source and target code, but this alignment is purely statement-level and solely relies on the LLM, without structural or semantic guidance~\cite{transMap}. As a result, this mapping approach is fragile in complex mapping scenarios; for example, a single source line may translate into multiple target lines, or lines may be shifted, ultimately compromising the effectiveness of error repair. 
To address these issues, we propose \ourtool{}. In the mapping phase, \ourtool{} employs CFG-based decomposition to partition the source program into atomic blocks, each forming an independent structural unit. LLMs then map these blocks to the target program. This hybrid approach improves mapping robustness and mitigates the limitations of \transmap{}.
In the fixing phase, \ourtool{} compares the intermediate execution states of the source and target programs to detect semantic discrepancies. This runtime-based localization offers clearer repair cues, substantially improving error correction accuracy and overcoming the limitations of \unitrans{}.

\subsection{Program Repair}
Program repair involves fixing bugs in the code. Early methods relied on heuristic approaches~\cite{DBLP:journals/tse/YuanB20}, constraint-based techniques~\cite{DBLP:journals/corr/abs-1811-04211}, and pattern-based methods~\cite{DBLP:journals/ese/KoyuncuLBKKMT20, DBLP:conf/wcre/LiuK0B19}. However, these approaches, which depend on manually designed templates, struggle with generalizing to diverse repair tasks. Learning-based methods like CoCoNut~\cite{DBLP:conf/issta/LutellierPPLW020} and Tare~\cite{DBLP:conf/icse/ZhuSZXZ23} improve performance by learning bug-fixing patterns from large datasets~\cite{DBLP:journals/tosem/ZhangFMSC24}. 
Recent LLM-based methods such as RING~\cite{DBLP:conf/aaai/JoshiSG0VR23}, \autocoder~\cite{AutoCodeRover}, and \agentless~\cite{Agentless} perform program repair using static information, such as issue descriptions and code similarity. However, these approaches are primarily designed for single-program scenarios within a single language, making them ill-suited for code translation tasks that involve reasoning across both source and target programs in different languages.
In contrast, \ourtool{} leverages dynamic information by 
analyzing the runtime behaviors of basic blocks partitioned via control-flow analysis in both the source and target programs. This allows for more precise fault localization and significantly improves repair performance. Our empirical results further demonstrate that \ourtool{} outperforms \agentless{} in the context of code translation error correction.

\section{Conclusion}
In this paper, we introduce \ourtool{}, a novel multi-agent system to improve LLM-based code translation with fine-grained execution alignment. 
The experimental results show that \ourtool{} outperforms the latest LLM-based code translation technique \unitrans{} in both translation effectiveness and efficiency; additionally, our ablation study demonstrates the contribution of each agent, and the error repair evaluation shows that \ourtool{} exhibits superior error-repair capabilities compared with \agentless{}; lastly, we further evaluate the generalization of \ourtool{} across different LLMs.

\section{Data Availability}
Our data and code are included in our replication package~\cite{TransAGENT}.

\begin{acks}
This work is supported by the National Key R\&D Program of China (Grant No.~2023YFB4503805) and the National Natural Science Foundation of China (Grant No.~62332005). 
\end{acks}

\balance
\bibliographystyle{ACM-Reference-Format}

\bibliography{ref}

@inproceedings{transMap,
  author       = {Bo Wang and
                  Ruishi Li and
                  Mingkai Li and
                  Prateek Saxena},
  title        = {TransMap: Pinpointing Mistakes in Neural Code Translation},
  booktitle    = {Proceedings of the 31st {ACM} Joint European Software Engineering
                  Conference and Symposium on the Foundations of Software Engineering,
                  {ESEC/FSE} 2023, San Francisco, CA, USA, December 3-9, 2023},
  pages        = {999--1011},
  publisher    = {{ACM}},
  year         = {2023}
}

@inproceedings{yang2025bamas,
  author       = {Yang, Liming and Luo, Junyu and Liu, Xuanzhe and Lou, Yiling and Chen, Zhenpeng},
  title        = {BAMAS: Structuring Budget-Aware Multi-Agent Systems},
  booktitle    = {Proceedings of the 40th Annual AAAI Conference on Artificial Intelligence, AAAI 2026},
  year         = {2026}
}

@inproceedings{DBLP:conf/nips/LuGRHSBCDJTLZSZ21,
  author       = {Shuai Lu and
                  Daya Guo and
                  Shuo Ren and
                  Junjie Huang and
                  Alexey Svyatkovskiy and
                  Ambrosio Blanco and
                  Colin B. Clement and
                  Dawn Drain and et al.},
  title        = {CodeXGLUE: {A} Machine Learning Benchmark Dataset for Code Understanding
                  and Generation},
  booktitle    = {Proceedings of the Neural Information Processing Systems Track on
                  Datasets and Benchmarks 1, NeurIPS Datasets and Benchmarks 2021, December
                  2021, virtual},
  year         = {2021},
}

@inproceedings{DBLP:conf/acl/AhmadTCC23,
  author       = {Wasi Uddin Ahmad and
                  Md Golam Rahman Tushar and
                  Saikat Chakraborty and
                  Kai{-}Wei Chang},
  title        = {{AVATAR:} {A} Parallel Corpus for Java-Python Program Translation},
  booktitle    = {Findings of the Association for Computational Linguistics: {ACL} 2023,
                  Toronto, Canada, July 9-14, 2023},
  pages        = {2268--2281},
  publisher    = {Association for Computational Linguistics},
  year         = {2023},
}

@inproceedings{DBLP:conf/sigsoft/NguyenNN13,
  author       = {Anh Tuan Nguyen and
                  Tung Thanh Nguyen and
                  Tien N. Nguyen},
  title        = {Lexical statistical machine translation for language migration},
  booktitle    = {Joint Meeting of the European Software Engineering Conference and
                  the {ACM} {SIGSOFT} Symposium on the Foundations of Software Engineering,
                  ESEC/FSE'13,},
  pages        = {651--654},
  publisher    = {{ACM}},
  year         = {2013},
  doi          = {10.1145/2491411.2494584},
}

@inproceedings{DBLP:conf/iclr/ChenLS18a,
  author       = {Xinyun Chen and
                  Chang Liu and
                  Dawn Song},
  title        = {Tree-to-tree Neural Networks for Program Translation},
  booktitle    = {6th International Conference on Learning Representations, {ICLR} 2018,
                  Vancouver, BC, Canada, April 30 - May 3, 2018, Workshop Track Proceedings},
  publisher    = {OpenReview.net},
  year         = {2018},
}

@inproceedings{DBLP:conf/nips/RoziereLCL20,
  author       = {Baptiste Rozi{\`{e}}re and
                  Marie{-}Anne Lachaux and
                  Lowik Chanussot and
                  Guillaume Lample},
  title        = {Unsupervised Translation of Programming Languages},
  booktitle    = {Advances in Neural Information Processing Systems 33: Annual Conference
                  on Neural Information Processing Systems 2020, NeurIPS 2020, December
                  6-12, 2020, virtual},
  year         = {2020},
}

@article{DBLP:journals/corr/abs-2105-12655,
  author       = {Ruchir Puri and
                  David S. Kung and
                  Geert Janssen and
                  Wei Zhang and
                  Giacomo Domeniconi and
                  Vladimir Zolotov and et al.
                  },
  title        = {Project CodeNet: {A} Large-Scale {AI} for Code Dataset for Learning
                  a Diversity of Coding Tasks},
  journal      = {CoRR},
  volume       = {abs/2105.12655},
  year         = {2021},
  eprinttype    = {arXiv},
  eprint       = {2105.12655}
}

@inproceedings{DBLP:conf/aaai/Zhu0R22,
  author       = {Ming Zhu and
                  Karthik Suresh and
                  Chandan K. Reddy},
  title        = {Multilingual Code Snippets Training for Program Translation},
  booktitle    = {Thirty-Sixth {AAAI} Conference on Artificial Intelligence, {AAAI}
                  2022, Thirty-Fourth Conference on Innovative Applications of Artificial
                  Intelligence, {IAAI} 2022, The Twelveth Symposium on Educational Advances
                  in Artificial Intelligence, {EAAI} 2022 Virtual Event, February 22
                  - March 1, 2022},
  pages        = {11783--11790},
  publisher    = {{AAAI} Press},
  year         = {2022},
  doi          = {10.1609/AAAI.V36I10.21434}
}

@article{DBLP:journals/corr/abs-2009-10297,
  author       = {Shuo Ren and
                  Daya Guo and
                  Shuai Lu and
                  Long Zhou and
                  Shujie Liu and
                  Duyu Tang and
                  Neel Sundaresan and
                  Ming Zhou and
                  Ambrosio Blanco and
                  Shuai Ma},
  title        = {CodeBLEU: a Method for Automatic Evaluation of Code Synthesis},
  journal      = {CoRR},
  volume       = {abs/2009.10297},
  year         = {2020},
  eprinttype    = {arXiv},
  eprint       = {2009.10297}
}

@article{minimumArrayLength,
  author={minimumArrayLength},
  year={2024.01},
  note = {\url{https://leetcode.cn/problems/minimize-length-of-array-using-operations/solutions/2613059/on-nao-jin-ji-zhuan-wan-pythonjavacgo-by-2lea}}
}

@article{minOperations,
  author={minOperations},
  year={2024.03},
  note = {\url{https://leetcode.cn/problems/minimum-operations-to-exceed-threshold-value-ii/solutions/2664316/zui-xiao-dui-mo-ni-pythonjavacgo-by-endl-07jk/}}
}

@article{gpt4omini,
  note = {\url{https://openai.com/index/gpt-4o-mini-advancing-cost-efficient-intelligence/}},
  author={gpt-4o-mini},
  year={2024}
}

@article{LlamaInstruct,
  note = {\url{https://huggingface.co/meta-llama/Meta-Llama-3-8B-Instruct}},
  author={Llama},
}

@article{huggingface,
  note = {\url{https://huggingface.co}},
  author={huggingface}
}

@article{joern,
  note = {\url{https://docs.joern.io/server/}},
  author={joern}
}

@article{ChatGLM2,
  note = {\url{https://github.com/THUDM/ChatGLM2-6B}},
  author={ChatGLM2},
}

@article{ChatGLM,
  note = {\url{https://github.com/THUDM/ChatGLM-6B}},
  author={ChatGLM},
}

@article{deepseek,
  note = {\url{https://huggingface.co/deepseek-ai/deepseek-coder-6.7b-instruct}},
  author={Deepseek},
}

@article{leetcode,
  note = {\url{https://leetcode.cn/}},
  author={leetcode}
}

@article{geeksforgeeks,
  note = {\url{https://www.geeksforgeeks.org/}},
  author={geeksforgeeks}
}

@incollection{DBLP:reference/stat/ReyN11,
  author       = {Denise Rey and
                  Markus Neuh{\"{a}}user},
  editor       = {Miodrag Lovric},
  title        = {Wilcoxon-Signed-Rank Test},
  booktitle    = {International Encyclopedia of Statistical Science},
  pages        = {1658--1659},
  publisher    = {Springer},
  year         = {2011},
  url          = {https://doi.org/10.1007/978-3-642-04898-2\_616},
  doi          = {10.1007/978-3-642-04898-2\_616},
  bibsource    = {dblp computer science bibliography, https://dblp.org}
}

@article{unitrans,
  author       = {Zhen Yang and
                  Fang Liu and
                  Zhongxing Yu and
                  Jacky Wai Keung and
                  Jia Li and
                  Shuo Liu and
                  Yifan Hong and
                  Xiaoxue Ma and
                  Zhi Jin and
                  Ge Li},
  title        = {Exploring and Unleashing the Power of Large Language Models in Automated
                  Code Translation},
  journal      = {Proc. {ACM} Softw. Eng.},
  volume       = {1},
  number       = {{FSE}},
  pages        = {1585--1608},
  year         = {2024},
  doi          = {10.1145/3660778}
}

@inproceedings{DBLP:conf/iclr/RoziereZCHSL22,
  author       = {Baptiste Rozi{\`{e}}re and
                  Jie Zhang and
                  Fran{\c{c}}ois Charton and
                  Mark Harman and
                  Gabriel Synnaeve and
                  Guillaume Lample},
  title        = {Leveraging Automated Unit Tests for Unsupervised Code Translation},
  booktitle    = {The Tenth International Conference on Learning Representations, {ICLR}
                  2022, Virtual Event, April 25-29, 2022},
  publisher    = {OpenReview.net}
}

@inproceedings{DBLP:conf/nips/Wei0SBIXCLZ22,
  author       = {Jason Wei and
                  Xuezhi Wang and
                  Dale Schuurmans and
                  Maarten Bosma and
                  Brian Ichter and
                  Fei Xia and
                  Ed H. Chi and
                  Quoc V. Le and
                  Denny Zhou},
  title        = {Chain-of-Thought Prompting Elicits Reasoning in Large Language Models},
  booktitle    = {Advances in Neural Information Processing Systems 35: Annual Conference
                  on Neural Information Processing Systems 2022, NeurIPS 2022, New Orleans,
                  LA, USA, November 28 - December 9, 2022},
  year         = {2022}
}

@inproceedings{DBLP:conf/iclr/0002WSLCNCZ23,
  author       = {Xuezhi Wang and
                  Jason Wei and
                  Dale Schuurmans and
                  Quoc V. Le and
                  Ed H. Chi and
                  Sharan Narang and
                  Aakanksha Chowdhery and
                  Denny Zhou},
  title        = {Self-Consistency Improves Chain of Thought Reasoning in Language Models},
  booktitle    = {The Eleventh International Conference on Learning Representations,
                  {ICLR} 2023, Kigali, Rwanda, May 1-5, 2023},
  publisher    = {OpenReview.net},
  year         = {2023}
}

@article{DBLP:journals/corr/abs-2310-04959,
  author       = {Zihan Yu and
                  Liang He and
                  Zhen Wu and
                  Xinyu Dai and
                  Jiajun Chen},
  title        = {Towards Better Chain-of-Thought Prompting Strategies: {A} Survey},
  journal      = {CoRR},
  volume       = {abs/2310.04959},
  year         = {2023},
  url          = {https://doi.org/10.48550/arXiv.2310.04959},
  doi          = {10.48550/ARXIV.2310.04959},
  eprinttype    = {arXiv},
  eprint       = {2310.04959}
}

@inproceedings{DBLP:conf/icse/PanIKSWMSPSJ24,
  author       = {Rangeet Pan and
                  Ali Reza Ibrahimzada and
                  Rahul Krishna and
                  Divya Sankar and
                  et al.},
  title        = {Lost in Translation: {A} Study of Bugs Introduced by Large Language
                  Models while Translating Code},
  booktitle    = {Proceedings of the 46th {IEEE/ACM} International Conference on Software
                  Engineering, {ICSE} 2024, Lisbon, Portugal, April 14-20, 2024},
  pages        = {82:1--82:13},
  publisher    = {{ACM}},
  year         = {2024}
}

@inproceedings{DBLP:conf/nips/LiuXW023,
  author       = {Jiawei Liu and
                  Chunqiu Steven Xia and
                  Yuyao Wang and
                  Lingming Zhang},
  title        = {Is Your Code Generated by ChatGPT Really Correct? Rigorous Evaluation
                  of Large Language Models for Code Generation},
  booktitle    = {Advances in Neural Information Processing Systems 36: Annual Conference
                  on Neural Information Processing Systems 2023, NeurIPS 2023, New Orleans,
                  LA, USA, December 10 - 16, 2023},
  year         = {2023},
  bibsource    = {dblp computer science bibliography, https://dblp.org}
}

@article{chen2024reasoning,
  title={Reasoning Runtime Behavior of a Program with LLM: How Far Are We?},
  author={Chen, Junkai and Pan, Zhiyuan and Hu, Xing and Li, Zhenhao and Li, Ge and Xia, Xin},
  journal={arXiv e-prints},
  year={2024}
}

@inproceedings{DBLP:conf/apsec/TaoYGS24,
  author       = {Qingxiao Tao and
                  Tingrui Yu and
                  Xiaodong Gu and
                  Beijun Shen},
  title        = {Unraveling the Potential of Large Language Models in Code Translation:
                  How Far are We?},
  booktitle    = {31st Asia-Pacific Software Engineering Conference, {APSEC} 2024, Chongqing,
                  China, December 3-6, 2024},
  pages        = {353--362},
  publisher    = {{IEEE}},
  year         = {2024},
  url          = {https://doi.org/10.1109/APSEC65559.2024.00046},
  doi          = {10.1109/APSEC65559.2024.00046},
  bibsource    = {dblp computer science bibliography, https://dblp.org}
}

@article{DBLP:journals/corr/abs-2412-14234,
  author       = {Manish Shetty and
                  Naman Jain and
                  Adwait Godbole and
                  Sanjit A. Seshia and
                  Koushik Sen},
  title        = {Syzygy: Dual Code-Test {C} to (safe) Rust Translation using LLMs and
                  Dynamic Analysis},
  journal      = {CoRR},
  volume       = {abs/2412.14234},
  year         = {2024},
  url          = {https://doi.org/10.48550/arXiv.2412.14234},
  doi          = {10.48550/ARXIV.2412.14234},
  eprinttype    = {arXiv},
  eprint       = {2412.14234},
  bibsource    = {dblp computer science bibliography, https://dblp.org}
}

@ARTICLE{7273801,
  author={Echeverria, Roberto Rodriguez and Macias, Fernando and Pavon, Victor Manuel and Conejero, Jose Maria and Figueroa, Fernando Sanchez},
  journal={IEEE Latin America Transactions}, 
  title={Legacy Web Application Modernization by Generating a REST Service Layer}, 
  year={2015},
  volume={13},
  number={7},
  pages={2379-2383},
  doi={10.1109/TLA.2015.7273801}}

@inproceedings{DBLP:conf/euromicro/HaugelandNSC21,
  author       = {Sindre Gr{\o}nst{\o}l Haugeland and
                  Phu Hong Nguyen and
                  Hui Song and
                  Franck Chauvel},
  title        = {Migrating Monoliths to Microservices-based Customizable Multi-tenant
                  Cloud-native Apps},
  booktitle    = {47th Euromicro Conference on Software Engineering and Advanced Applications,
                  {SEAA} 2021, Palermo, Italy, September 1-3, 2021},
  pages        = {170--177},
  publisher    = {{IEEE}},
  year         = {2021}
}

@INPROCEEDINGS{9678878,
  author={},
  booktitle={2021 36th IEEE/ACM International Conference on Automated Software Engineering (ASE)}, 
  title={Transforming Monolithic Applications to Microservices with Mono2Micro}, 
  year={2021},
  volume={},
  number={},
  pages={3-3},
  doi={10.1109/ASE51524.2021.9678878}
}

@article{DBLP:journals/corr/abs-2004-10724,
  author       = {Mahdi Fahmideh and
                  Farhad Daneshgar and
                  Ghassan Beydoun and
                  Fethi A. Rabhi},
  title        = {Challenges in migrating legacy software systems to the cloud an empirical
                  study},
  journal      = {CoRR},
  volume       = {abs/2004.10724},
  year         = {2020},
  eprinttype    = {arXiv},
  eprint       = {2004.10724}
}

@inproceedings{DBLP:conf/eacl/AhmadCRC23,
  author       = {Wasi Uddin Ahmad and
                  Saikat Chakraborty and
                  Baishakhi Ray and
                  Kai{-}Wei Chang},
  title        = {Summarize and Generate to Back-translate: Unsupervised Translation
                  of Programming Languages},
  booktitle    = {Proceedings of the 17th Conference of the European Chapter of the
                  Association for Computational Linguistics, {EACL} 2023, Dubrovnik,
                  Croatia, May 2-6, 2023},
  pages        = {1520--1534},
  publisher    = {Association for Computational Linguistics},
  year         = {2023}
}

@inproceedings{DBLP:conf/emnlp/XieNFR23,
  author       = {Yiqing Xie and
                  Atharva Naik and
                  Daniel Fried and
                  Carolyn P. Ros{\'{e}}},
  title        = {Data Augmentation for Code Translation with Comparable Corpora and
                  Multiple References},
  booktitle    = {Findings of the Association for Computational Linguistics: {EMNLP}
                  2023, Singapore, December 6-10, 2023},
  pages        = {13725--13739},
  publisher    = {Association for Computational Linguistics},
  year         = {2023}
}

@inproceedings{DBLP:conf/icse/FanGMRT23,
  author       = {Zhiyu Fan and
                  Xiang Gao and
                  Martin Mirchev and
                  Abhik Roychoudhury and
                  Shin Hwei Tan},
  title        = {Automated Repair of Programs from Large Language Models},
  booktitle    = {45th {IEEE/ACM} International Conference on Software Engineering,
                  {ICSE} 2023, Melbourne, Australia, May 14-20, 2023},
  pages        = {1469--1481},

  year         = {2023},
  doi          = {10.1109/ICSE48619.2023.00128}
}

@inproceedings{DBLP:conf/icse/XiaWZ23,
  author       = {Chunqiu Steven Xia and
                  Yuxiang Wei and
                  Lingming Zhang},
  title        = {Automated Program Repair in the Era of Large Pre-trained Language
                  Models},
  booktitle    = {45th {IEEE/ACM} International Conference on Software Engineering,
                  {ICSE} 2023, Melbourne, Australia, May 14-20, 2023},
  pages        = {1482--1494},
  publisher    = {{IEEE}},
  year         = {2023},
  doi          = {10.1109/ICSE48619.2023.00129}
}

@article{DBLP:journals/corr/abs-2304-11384,
  author       = {Mingyang Geng and
                  Shangwen Wang and
                  Dezun Dong and
                  Haotian Wang and
                  Ge Li and
                  Zhi Jin and
                  Xiaoguang Mao and
                  Xiangke Liao},
  title        = {An Empirical Study on Using Large Language Models for Multi-Intent
                  Comment Generation},
  journal      = {CoRR},
  volume       = {abs/2304.11384},
  year         = {2023},
  doi          = {10.48550/ARXIV.2304.11384},
  eprinttype    = {arXiv},
  eprint       = {2304.11384}
}

@inproceedings{DBLP:conf/kbse/AhmedD22,
  author       = {Toufique Ahmed and
                  Premkumar T. Devanbu},
  title        = {Few-shot training LLMs for project-specific code-summarization},
  booktitle    = {37th {IEEE/ACM} International Conference on Automated Software Engineering,
                  {ASE} 2022, Rochester, MI, USA, October 10-14, 2022},
  pages        = {177:1--177:5},
  publisher    = {{ACM}},
  year         = {2022},
  doi          = {10.1145/3551349.3559555}
}

@article{c2rust,
  note = {/url{https://github.com/immunant/c2rust}},
  author={C2Rust},
  year = {2024}, 
}

@article{cxgo,
  note = {\url{https://github.com/gotranspile/cxgo}},
  author={cxgo: C to Go transpiler},
  year = {2024}, 
}

@misc{sharpen,
  author = {Sharpen},
  year = {2020},
  howpublished = {\url{https://github.com/mono/sharpen}},
}

@misc{JavaToCSharp,
  author = {JavaToCSharp},
  year = {2024}, 
  howpublished = {\url{https://github.com/paulirwin/JavaToCSharp}},
  
}

@inproceedings{DBLP:conf/oopsla/KaraivanovRV14,
  author       = {Svetoslav Karaivanov and
                  Veselin Raychev and
                  Martin T. Vechev},
  title        = {Phrase-Based Statistical Translation of Programming Languages},
  booktitle    = {Onward! 2014, Proceedings of the 2014 {ACM} International Symposium
                  on New Ideas, New Paradigms, and Reflections on Programming {\&}
                  Software, part of {SPLASH} '14, Portland, OR, USA, October 20-24,
                  2014},
  pages        = {173--184}
}

@inproceedings{DBLP:conf/nips/LachauxRSL21,
  author       = {Marie{-}Anne Lachaux and
                  Baptiste Rozi{\`{e}}re and
                  Marc Szafraniec and
                  Guillaume Lample},
  title        = {{DOBF:} {A} Deobfuscation Pre-Training Objective for Programming Languages},
  booktitle    = {Advances in Neural Information Processing Systems 34: Annual Conference
                  on Neural Information Processing Systems 2021, NeurIPS 2021, December
                  6-14, 2021, virtual},
  pages        = {14967--14979},
  year         = {2021}
}

@inproceedings{DBLP:conf/kbse/OdaFNHSTN15,
  author       = {Yusuke Oda and
                  Hiroyuki Fudaba and
                  Graham Neubig and
                  Hideaki Hata and
                  Sakriani Sakti and
                  Tomoki Toda and
                  Satoshi Nakamura},
  title        = {Learning to Generate Pseudo-Code from Source Code Using Statistical
                  Machine Translation {(T)}},
  booktitle    = {30th {IEEE/ACM} International Conference on Automated Software Engineering,
                  {ASE} 2015},
  pages        = {574--584},
  publisher    = {{IEEE} Computer Society},
  year         = {2015},
  doi          = {10.1109/ASE.2015.36},
  bibsource    = {dblp computer science bibliography, https://dblp.org}
}

@inproceedings{DBLP:conf/iclr/SzafraniecRLLCS23,
  author       = {Marc Szafraniec and
                  Baptiste Rozi{\`{e}}re and
                  Hugh Leather and
                  Patrick Labatut and
                  Fran{\c{c}}ois Charton and
                  Gabriel Synnaeve},
  title        = {Code Translation with Compiler Representations},
  booktitle    = {The Eleventh International Conference on Learning Representations,
                  {ICLR} 2023, Kigali, Rwanda, May 1-5, 2023},
  publisher    = {OpenReview.net},
  year         = {2023},
  bibsource    = {dblp computer science bibliography, https://dblp.org}
}

@article{DBLP:journals/tse/YuanB20,
  author       = {Yuan Yuan and
                  Wolfgang Banzhaf},
  title        = {{ARJA:} Automated Repair of Java Programs via Multi-Objective Genetic
                  Programming},
  journal      = {{IEEE} Trans. Software Eng.},
  volume       = {46},
  number       = {10},
  pages        = {1040--1067},
  year         = {2020},
  doi          = {10.1109/TSE.2018.2874648},
  bibsource    = {dblp computer science bibliography, https://dblp.org}
}

@article{DBLP:journals/corr/abs-1811-04211,
  author       = {Jifeng Xuan and
                  Matias Martinez and
                  Favio Demarco and
                  Maxime Cl{\'{e}}ment and et al.},
  title        = {Nopol: Automatic Repair of Conditional Statement Bugs in Java Programs},
  journal      = {CoRR},
  volume       = {abs/1811.04211},
  year         = {2018},
  eprinttype    = {arXiv},
  eprint       = {1811.04211},
  bibsource    = {dblp computer science bibliography, https://dblp.org}
}

@article{DBLP:journals/ese/KoyuncuLBKKMT20,
  author       = {Anil Koyuncu and
                  Kui Liu and
                  Tegawend{\'{e}} F. Bissyand{\'{e}} and
                  Dongsun Kim and
                  Jacques Klein and
                  Martin Monperrus and
                  Yves Le Traon},
  title        = {FixMiner: Mining relevant fix patterns for automated program repair},
  journal      = {Empir. Softw. Eng.},
  volume       = {25},
  number       = {3},
  pages        = {1980--2024},
  year         = {2020},
  bibsource    = {dblp computer science bibliography, https://dblp.org}
}

@inproceedings{DBLP:conf/wcre/LiuK0B19,
  author       = {Kui Liu and
                  Anil Koyuncu and
                  Dongsun Kim and
                  Tegawend{\'{e}} F. Bissyand{\'{e}}},
  title        = {{AVATAR:} Fixing Semantic Bugs with Fix Patterns of Static Analysis
                  Violations},
  booktitle    = {26th {IEEE} International Conference on Software Analysis, Evolution
                  and Reengineering, {SANER} 2019, Hangzhou, China, February 24-27,
                  2019},
  pages        = {456--467},
  publisher    = {{IEEE}},
  year         = {2019},
  bibsource    = {dblp computer science bibliography, https://dblp.org}
}

@inproceedings{DBLP:conf/issta/LutellierPPLW020,
  author       = {Thibaud Lutellier and
                  Hung Viet Pham and
                  Lawrence Pang and
                  Yitong Li and
                  Moshi Wei and
                  Lin Tan},
  title        = {CoCoNuT: combining context-aware neural translation models using ensemble
                  for program repair},
  booktitle    = {{ISSTA} '20: 29th {ACM} {SIGSOFT} International Symposium on Software
                  Testing and Analysis, Virtual Event, USA, July 18-22, 2020},
  pages        = {101--114},
  publisher    = {{ACM}},
  year         = {2020},
  bibsource    = {dblp computer science bibliography, https://dblp.org}
}

@inproceedings{DBLP:conf/icse/ZhuSZXZ23,
  author       = {Qihao Zhu and
                  Zeyu Sun and
                  Wenjie Zhang and
                  Yingfei Xiong and
                  Lu Zhang},
  title        = {Tare: Type-Aware Neural Program Repair},
  booktitle    = {45th {IEEE/ACM} International Conference on Software Engineering,
                  {ICSE} 2023, Melbourne, Australia, May 14-20, 2023},
  pages        = {1443--1455},
  publisher    = {{IEEE}},
  year         = {2023},
  doi          = {10.1109/ICSE48619.2023.00126},
  bibsource    = {dblp computer science bibliography, https://dblp.org}
}

@article{DBLP:journals/tosem/ZhangFMSC24,
  author       = {Quanjun Zhang and
                  Chunrong Fang and
                  Yuxiang Ma and
                  Weisong Sun and
                  Zhenyu Chen},
  title        = {A Survey of Learning-based Automated Program Repair},
  journal      = {{ACM} Trans. Softw. Eng. Methodol.},
  volume       = {33},
  number       = {2},
  pages        = {55:1--55:69},
  year         = {2024},
  doi          = {10.1145/3631974},
  bibsource    = {dblp computer science bibliography, https://dblp.org}
}

@inproceedings{DBLP:conf/kbse/ZhangFZYSC23,
  author       = {Quanjun Zhang and
                  Chunrong Fang and
                  Tongke Zhang and
                  Bowen Yu and
                  Weisong Sun and
                  Zhenyu Chen},
  title        = {Gamma: Revisiting Template-Based Automated Program Repair Via Mask
                  Prediction},
  booktitle    = {38th {IEEE/ACM} International Conference on Automated Software Engineering,
                  {ASE} 2023, Luxembourg, September 11-15, 2023},
  pages        = {535--547},
  publisher    = {{IEEE}},
  year         = {2023},
  bibsource    = {dblp computer science bibliography, https://dblp.org}
}

@inproceedings{DBLP:conf/aaai/JoshiSG0VR23,
  author       = {Harshit Joshi and
                  Jos{\'{e}} Pablo Cambronero S{\'{a}}nchez and
                  Sumit Gulwani and
                  Vu Le and
                  Gust Verbruggen and
                  Ivan Radicek},
  title        = {Repair Is Nearly Generation: Multilingual Program Repair with LLMs},
  booktitle    = {Thirty-Seventh {AAAI} Conference on Artificial Intelligence, {AAAI}
                  2023, Thirty-Fifth Conference on Innovative Applications of Artificial
                  Intelligence, {IAAI} 2023, Thirteenth Symposium on Educational Advances
                  in Artificial Intelligence, {EAAI} 2023, Washington, DC, USA, February
                  7-14, 2023},
  pages        = {5131--5140},
  publisher    = {{AAAI} Press},
  year         = {2023},
  doi          = {10.1609/AAAI.V37I4.25642},
  bibsource    = {dblp computer science bibliography, https://dblp.org}
}

@inproceedings{DBLP:conf/sigsoft/XiaZ22,
  author       = {Chunqiu Steven Xia and
                  Lingming Zhang},
  title        = {Less training, more repairing please: revisiting automated program
                  repair via zero-shot learning},
  booktitle    = {Proceedings of the 30th {ACM} Joint European Software Engineering
                  Conference and Symposium on the Foundations of Software Engineering,
                  {ESEC/FSE} 2022, Singapore, Singapore, November 14-18, 2022},
  pages        = {959--971},
  publisher    = {{ACM}},
  year         = {2022},
  doi          = {10.1145/3540250.3549101},
  bibsource    = {dblp computer science bibliography, https://dblp.org}
}

@inproceedings{DBLP:conf/kbse/XiaDZ23,
  author       = {Chunqiu Steven Xia and
                  Yifeng Ding and
                  Lingming Zhang},
  title        = {The Plastic Surgery Hypothesis in the Era of Large Language Models},
  booktitle    = {38th {IEEE/ACM} International Conference on Automated Software Engineering,
                  {ASE} 2023, Luxembourg, September 11-15, 2023},
  pages        = {522--534},
  publisher    = {{IEEE}},
  year         = {2023},
  doi          = {10.1109/ASE56229.2023.00047},
  bibsource    = {dblp computer science bibliography, https://dblp.org}
}

@inproceedings{DBLP:conf/sigsoft/0003X023,
  author       = {Yuxiang Wei and
                  Chunqiu Steven Xia and
                  Lingming Zhang},
  title        = {Copiloting the Copilots: Fusing Large Language Models with Completion
                  Engines for Automated Program Repair},
  booktitle    = {Proceedings of the 31st {ACM} Joint European Software Engineering
                  Conference and Symposium on the Foundations of Software Engineering,
                  {ESEC/FSE} 2023, San Francisco, CA, USA, December 3-9, 2023},
  pages        = {172--184},
  publisher    = {{ACM}},
  year         = {2023},
  doi          = {10.1145/3611643.3616271},
  bibsource    = {dblp computer science bibliography, https://dblp.org}
}

@article{TransAGENT,
  url = {https://github.com/FudanSELab/TransAgent},
  author = {TransAGENT},
  year = {2025},

}

@INPROCEEDINGS{953283,
  author={Sherwood, T. and Perelman, E. and Calder, B.},
  booktitle={Proceedings 2001 International Conference on Parallel Architectures and Compilation Techniques}, 
  title={Basic block distribution analysis to find periodic behavior and simulation points in applications}, 
  year={2001},
  volume={},
  number={},
  pages={3-14},
  doi={10.1109/PACT.2001.953283}}

@inproceedings{AutoCodeRover,
  author       = {Yuntong Zhang and
                  Haifeng Ruan and
                  Zhiyu Fan and
                  Abhik Roychoudhury},
  title        = {AutoCodeRover: Autonomous Program Improvement},
  booktitle    = {Proceedings of the 33rd {ACM} {SIGSOFT} International Symposium on
                  Software Testing and Analysis, {ISSTA} 2024, Vienna, Austria, September
                  16-20, 2024},
  pages        = {1592--1604},
  publisher    = {{ACM}},
  year         = {2024},
  doi          = {10.1145/3650212.3680384},
  bibsource    = {dblp computer science bibliography, https://dblp.org}
}

@article{Agentless,
  author       = {Chunqiu Steven Xia and
                  Yinlin Deng and
                  Soren Dunn and
                  Lingming Zhang},
  title        = {Agentless: Demystifying LLM-based Software Engineering Agents},
  journal      = {CoRR},
  volume       = {abs/2407.01489},
  year         = {2024},
  url          = {https://doi.org/10.48550/arXiv.2407.01489},
  doi          = {10.48550/ARXIV.2407.01489},
  eprinttype    = {arXiv},
  eprint       = {2407.01489},
  bibsource    = {dblp computer science bibliography, https://dblp.org}
}

@article{DBLP:journals/corr/abs-2503-21710,
  author       = {Boyang Yang and
                  Haoye Tian and
                  Jiadong Ren and
                  Shunfu Jin and
                  Yang Liu and
                  Feng Liu and
                  Bach Le},
  title        = {Enhancing Repository-Level Software Repair via Repository-Aware Knowledge
                  Graphs},
  journal      = {CoRR},
  volume       = {abs/2503.21710},
  year         = {2025},
  doi          = {10.48550/ARXIV.2503.21710},
  eprinttype    = {arXiv},
  eprint       = {2503.21710},
  bibsource    = {dblp computer science bibliography, https://dblp.org}
}

@article{DBLP:journals/corr/abs-2409-00899,
  author       = {Yizhou Liu and
                  Pengfei Gao and
                  Xinchen Wang and
                  Jie Liu and
                  Yexuan Shi and
                  Zhao Zhang and
                  Chao Peng},
  title        = {MarsCode Agent: AI-native Automated Bug Fixing},
  journal      = {CoRR},
  volume       = {abs/2409.00899},
  year         = {2024},
  doi          = {10.48550/ARXIV.2409.00899},
  eprinttype    = {arXiv},
  eprint       = {2409.00899},
  bibsource    = {dblp computer science bibliography, https://dblp.org}
}

@inproceedings{DBLP:conf/issta/JinO13,
  author       = {Wei Jin and
                  Alessandro Orso},
  title        = {{F3:} fault localization for field failures},
  booktitle    = {International Symposium on Software Testing and Analysis, {ISSTA}
                  '13, Lugano, Switzerland, July 15-20, 2013},
  pages        = {213--223},
  publisher    = {{ACM}},
  year         = {2013},
  url          = {https://doi.org/10.1145/2483760.2483763},
  doi          = {10.1145/2483760.2483763},
  bibsource    = {dblp computer science bibliography, https://dblp.org}
}

@inproceedings{DBLP:conf/uss/BlazytkoSAAFWH20,
  author       = {Tim Blazytko and
                  Moritz Schl{\"{o}}gel and
                  Cornelius Aschermann and
                  Ali Abbasi and
                  Joel Frank and
                  Simon W{\"{o}}rner and
                  Thorsten Holz},
  title        = {{AURORA:} Statistical Crash Analysis for Automated Root Cause Explanation},
  booktitle    = {29th {USENIX} Security Symposium, {USENIX} Security 2020, August 12-14,
                  2020},
  pages        = {235--252},
  publisher    = {{USENIX} Association},
  year         = {2020},
  url          = {https://www.usenix.org/conference/usenixsecurity20/presentation/blazytko},
  bibsource    = {dblp computer science bibliography, https://dblp.org}
}

@inproceedings{DBLP:conf/aaai/GuptaPKS17,
  author       = {Rahul Gupta and
                  Soham Pal and
                  Aditya Kanade and
                  Shirish K. Shevade},
  title        = {DeepFix: Fixing Common {C} Language Errors by Deep Learning},
  booktitle    = {Proceedings of the Thirty-First {AAAI} Conference on Artificial Intelligence,
                  February 4-9, 2017, San Francisco, California, {USA}},
  pages        = {1345--1351},
  publisher    = {{AAAI} Press},
  year         = {2017},
  doi          = {10.1609/AAAI.V31I1.10742},
  bibsource    = {dblp computer science bibliography, https://dblp.org}
}

@inproceedings{DBLP:conf/icse/LiuLZ23,
  author       = {Fang Liu and
                  Jia Li and
                  Li Zhang},
  title        = {Syntax and Domain Aware Model for Unsupervised Program Translation},
  booktitle    = {45th {IEEE/ACM} International Conference on Software Engineering,
                  {ICSE} 2023, Melbourne, Australia, May 14-20, 2023},
  pages        = {755--767},
  publisher    = {{IEEE}},
  year         = {2023},
  url          = {https://doi.org/10.1109/ICSE48619.2023.00072},
  doi          = {10.1109/ICSE48619.2023.00072},
  bibsource    = {dblp computer science bibliography, https://dblp.org}
}

@misc{tiobe,
  author = {{TIOBE Software}},
  title = {{TIOBE Programming Community Index}},
  year = {2025},
  url = {https://www.tiobe.com/tiobe-index},
}

@inproceedings{DBLP:conf/icse/WangLPLL25,
  author       = {Chong Wang and
                  Jianan Liu and
                  Xin Peng and
                  Yang Liu and
                  Yiling Lou},
  title        = {Boosting Static Resource Leak Detection via LLM-based Resource-Oriented
                  Intention Inference},
  booktitle    = {47th {IEEE/ACM} International Conference on Software Engineering,
                  {ICSE} 2025, Ottawa, ON, Canada, April 26 - May 6, 2025},
  pages        = {2905--2917},
  publisher    = {{IEEE}},
  year         = {2025},
  url          = {https://doi.org/10.1109/ICSE55347.2025.00131},
  doi          = {10.1109/ICSE55347.2025.00131},
  bibsource    = {dblp computer science bibliography, https://dblp.org}
}

@inproceedings{DBLP:conf/wcre/ChenGZ0WX24,
  author       = {Yujia Chen and
                  Cuiyun Gao and
                  Muyijie Zhu and
                  Qing Liao and
                  Yong Wang and
                  Guoai Xu},
  title        = {APIGen: Generative {API} Method Recommendation},
  booktitle    = {{IEEE} International Conference on Software Analysis, Evolution and
                  Reengineering, {SANER} 2024, Rovaniemi, Finland, March 12-15, 2024},
  pages        = {171--182},
  publisher    = {{IEEE}},
  year         = {2024},
  url          = {https://doi.org/10.1109/SANER60148.2024.00025},
  doi          = {10.1109/SANER60148.2024.00025},
  bibsource    = {dblp computer science bibliography, https://dblp.org}
}

@article{DBLP:journals/corr/abs-2409-15228,
  author       = {Yixi Wu and
                  Pengfei He and
                  Zehao Wang and
                  Shaowei Wang and
                  Yuan Tian and
                  Tse{-}Hsun Chen},
  title        = {A Comprehensive Framework for Evaluating API-oriented Code Generation
                  in Large Language Models},
  journal      = {CoRR},
  volume       = {abs/2409.15228},
  year         = {2024},
  url          = {https://doi.org/10.48550/arXiv.2409.15228},
  doi          = {10.48550/ARXIV.2409.15228},
  eprinttype    = {arXiv},
  eprint       = {2409.15228},
  bibsource    = {dblp computer science bibliography, https://dblp.org}
}

@article{DBLP:journals/pacmse/IbrahimzadaKPAPSJ25,
  author       = {Ali Reza Ibrahimzada and
                  Kaiyao Ke and
                  Mrigank Pawagi and
                  Muhammad Salman Abid and
                  Rangeet Pan and
                  Saurabh Sinha and
                  Reyhaneh Jabbarvand},
  title        = {AlphaTrans: {A} Neuro-Symbolic Compositional Approach for Repository-Level
                  Code Translation and Validation},
  journal      = {Proc. {ACM} Softw. Eng.},
  volume       = {2},
  number       = {{FSE}},
  pages        = {2454--2476},
  year         = {2025},
  url          = {https://doi.org/10.1145/3729379},
  doi          = {10.1145/3729379},
  bibsource    = {dblp computer science bibliography, https://dblp.org}
}

@misc{guan2025repotransagentmultiagentllmframework,
      title={RepoTransAgent: Multi-Agent LLM Framework for Repository-Aware Code Translation}, 
      author={Ziqi Guan and Xin Yin and Zhiyuan Peng and Chao Ni},
      year={2025},
      eprint={2508.17720},
      archivePrefix={arXiv},
      primaryClass={cs.SE},
      url={https://arxiv.org/abs/2508.17720}, 
}

@article{DBLP:journals/corr/abs-2503-17741,
  author       = {Xuemeng Cai and
                  Jiakun Liu and
                  Xiping Huang and
                  Yijun Yu and
                  Haitao Wu and
                  Chunmiao Li and
                  Bo Wang and
                  Imam Nur Bani Yusuf and
                  Lingxiao Jiang},
  title        = {RustMap: Towards Project-Scale C-to-Rust Migration via Program Analysis
                  and {LLM}},
  journal      = {CoRR},
  volume       = {abs/2503.17741},
  year         = {2025},
  url          = {https://doi.org/10.48550/arXiv.2503.17741},
  doi          = {10.48550/ARXIV.2503.17741},
  eprinttype    = {arXiv},
  eprint       = {2503.17741},
  bibsource    = {dblp computer science bibliography, https://dblp.org}
}

@inproceedings{LiuCLZHH0D0025,
  author       = {Kaibo Liu and
                  Zhenpeng Chen and
                  Yiyang Liu and
                  Jie M. Zhang and
                  Mark Harman and
                  Yudong Han and
                  Yun Ma and
                  Yihong Dong and
                  Ge Li and
                  Gang Huang},
  title        = {LLM-Powered Test Case Generation for Detecting Bugs in Plausible Programs},
  booktitle    = {Proceedings of the 63rd Annual Meeting of the Association for Computational
                  Linguistics (Volume 1: Long Papers), {ACL} 2025},
  pages        = {430--440},
  year         = {2025}
}

@inproceedings{rahardja2025can,
  author       = {Rahardja, Alfin Wijaya and Liu, Junwei and Chen, Weitong and Chen, Zhenpeng and Lou, Yiling},
  title        = {Can Agents Fix Agent Issues?},
  booktitle    = {Advances in Neural Information Processing Systems 39: Annual Conference
                  on Neural Information Processing Systems 2025, NeurIPS 2025},
  year         = {2025}
}

@article{abs240902977,
  author       = {Junwei Liu and
                  Kaixin Wang and
                  Yixuan Chen and
                  Xin Peng and
                  Zhenpeng Chen and
                  Lingming Zhang and
                  Yiling Lou},
  title        = {Large Language Model-Based Agents for Software Engineering: {A} Survey},
  journal      = {ACM Transactions on Software Engineering and Methodology},
  year         = {2026}
}

@article{guo2026eet,
  title={EET: Experience-Driven Early Termination for Cost-Efficient Software Engineering Agents},
  author={Guo, Yaoqi and Xiao, Ying and Zhang, Jie M and Harman, Mark and Lou, Yiling and Liu, Yang and Chen, Zhenpeng},
  journal={arXiv preprint arXiv:2601.05777},
  year={2026}
}

@article{DBLP:journals/tosem/HuangYXPXL24,
  author       = {Qing Huang and
                  Zhiqiang Yuan and
                  Zhenchang Xing and
                  Xin Peng and
                  Xiwei Xu and
                  Qinghua Lu},
  title        = {{FQN} Inference in Partial Code by Prompt-tuned Language Model of
                  Code},
  journal      = {{ACM} Trans. Softw. Eng. Methodol.},
  volume       = {33},
  number       = {2},
  pages        = {31:1--31:32},
  year         = {2024},
  url          = {https://doi.org/10.1145/3617174},
  doi          = {10.1145/3617174},
  bibsource    = {dblp computer science bibliography, https://dblp.org}
}

@article{DBLP:journals/tosem/RamosLMMG24,
  author       = {Daniel Ramos and
                  In{\^{e}}s Lynce and
                  Vasco Manquinho and
                  Ruben Martins and
                  Claire {Le Goues}},
  title        = {BatFix: Repairing language model-based transpilation},
  journal      = {{ACM} Trans. Softw. Eng. Methodol.},
  volume       = {33},
  number       = {6},
  pages        = {161},
  year         = {2024},
  url          = {https://doi.org/10.1145/3658668},
  doi          = {10.1145/3658668},
  bibsource    = {dblp computer science bibliography, https://dblp.org}
}



\end{document}